\documentstyle[12pt]{article}
\def\thebibliography#1{\bigskip\section*{\centering
References\\}\bigskip\list
{\arabic{enumi}.}{\settowidth\labelwidth{#1}\leftmargin\labelwidth
\advance\leftmargin\labelsep
\usecounter{enumi}}
\def\newblock{\hskip .11em plus .33em minus .07em}
\sloppy\clubpenalty4000\widowpenalty4000
\sfcode`\.=1000\relax}

\def\op#1{\mathop{\fam0 #1}\limits}
\newcommand{\Id}{{\rm Id\,}}
\def\Ker{{\rm Ker\,}}

\newcommand{\ben}{\begin{eqnarray}}
\newcommand{\een}{\end{eqnarray}}

\newcommand{\be}{\begin{eqnarray*}}
\newcommand{\ee}{\end{eqnarray*}}
\newcommand{\bea}{\begin{eqalph}}
\newcommand{\eea}{\end{eqalph}}
\newcommand{\cL}{{\cal L}}
\newcommand{\bL}{{\bf L}}
\newcommand{\R}{{\bf R}}
\newcommand{\cE}{{\cal E}}

\newcommand{\cF}{{\cal F}}
\newcommand{\cG}{{\cal G}}
\newcommand{\cT}{{\cal T}}

\newcommand{\al}{\alpha}
\newcommand{\bt}{\beta}

\newcommand{\dl}{\delta}
\newcommand{\la}{\lambda}

\newcommand{\ap}{\approx}

\newcommand{\f}{\phi}
\newcommand{\om}{\omega}
\newcommand{\ot}{\otimes}
\newcommand{\Om}{\Omega}
\newcommand{\m}{\mu}
\newcommand{\n}{\nu}
\newcommand{\g}{\gamma}
\newcommand{\G}{\Gamma}
\newcommand{\e}{\epsilon}
\newcommand{\ve}{\varepsilon}
\newcommand{\th}{\theta}
\newcommand{\r}{\rho}
\newcommand{\Si}{\Sigma}
\newcommand{\si}{\sigma}

\newcommand{\w}{\wedge}

\newcommand{\wt}{\widetilde}
\newcommand{\wh}{\widehat}
\newcommand{\ol}{\overline}
\newcommand{\dr}{\partial}

\newcounter{eqalph}
\newcounter{equationa}

\newenvironment{eqalph}{\stepcounter{equation}
\setcounter{equationa}{\value{equation}}
\setcounter{equation}{0}

\begin{eqnarray}}{\end{eqnarray}
\setcounter{equation}{\value{equationa}}}

\begin{document}
\hbox{}

\centerline{\large\bf Stress-Energy-Momentum Tensors}
\medskip

\centerline{\large\bf in Lagrangian Field Theory.}
\bigskip

\centerline{\large\bf Part 1. Superpotentials.}
\bigskip

\centerline{\sc Giovanni Giachetta}

\medskip

\centerline{Department of Mathematics and Physics}

\centerline{University of Camerino, 62032 Camerino, Italy}

\centerline{E-mail: mangiarotti@camvax.unicam.it}
\medskip

\centerline{\sc Gennadi Sardanashvily}
\medskip

\centerline{Department of Theoretical Physics}

\centerline{Moscow State University, 117234 Moscow, Russia}

\centerline{E-mail: sard@grav.phys.msu.su}
\bigskip

\begin{abstract}

Differential conservation laws in Lagrangian field theory are usually related
to symmetries of a Lagrangian density and are obtained if the Lie
derivative of a Lagrangian density by a certain class of vector fields on a
fiber bundle vanishes. However, only two field models meet this
property in fact. In gauge theory of exact internal symmetries, the Lie
derivative by vertical vector fields corresponding to gauge transformations
is equal to zero. The corresponding N\"oether current is reduced
to a superpotential that provides invariance of the N\"oether conservation law
under gauge transformations. In the gravitation theory, we
meet the phenomenon of "hidden energy".
Only the superpotential part of energy-momentum of gravity and matter is
observed when the general covariant transformations are exact.
Other parts of energy-momentum  display themselves if the
invariance under general covariance transformations is broken, e.g., by a
background world metric.
In this case, the Lie derivatives of
Lagrangian densities by vector fields which call into play the
stress-energy-momentum tensors fail to be equal to zero in general.  We base
our analysis of differential conservation laws on the canonical decomposition
of the Lie derivative of a Lagrangian density $L$ by a
 projectable vector field on a bundle and with respect to
different Lepagian equivalents of $L$. Different Lepagian
equivalents lead to conserved quantities which differ from each other in
superpotential terms. We have different stress-energy-momentum tensors
depending on different lifts of vector fields on a base onto a bundle.
Moreover, different solutions of the same Euler-Lagrange equations may require
different energy-momentum tensors.
We show that different stress-energy-momentum tensors
differ from each other in N\"oether currents.  As a consequence, the
energy-momentum conservation law can not take place if
internal symmetries are broken.
\end{abstract}
\newpage

{\bf CONTENTS}
\bigskip

\noindent
1. Discussion \newline
2. Geometric Preliminary
\medskip

{\sc Part 1.}
\newline
3. Lagrangian Formalism of Field Theory\newline
4. Conservation Laws\newline
5. Stress-Energy-Momentum Conservation Laws\newline
6. Stress-Energy Conservation Laws in Mechanics \newline
7. N\"oether Conservation Laws \newline
8. General Covariance Condition\newline
9. Stress-Energy-Momentum Tensor of Matter Fields \newline
10. Stress-Energy-Momentum Tensors of Gauge Potentials \newline
11. Stress-Energy-Momentum Tensors of Proca Fields \newline
12. Topological Gauge Theories
\medskip

{\sc Part 2.}
\newline
13. Reduced Second Order Lagrangian Formalism\newline
14. Conservation Laws in Einstein's Gravitation Theory\newline
15. First Order Palatini Formalism\newline
16. Stress-Energy-Momentum Tensors of Affine-Metric Gravity\newline
17. Lagrangian Systems on Composite Bundles\newline
18. Composite Spinor Bundles in Gravitation Theory\newline
19. Gauge Gravitation Theory\newline
20. Conservation Laws in the Gauge Gravitation Theory
\newpage

\section{Discussion}

The present work is devoted to
differential conservation laws in Lagrangian field theory.

We follow the geometric approach to field theory when classical fields are
described by global sections of a bundle $Y\to X$ over a world manifold $X$.
Their dynamics is phrased in terms of jet manifolds
\cite{ech1,kol,sard,sard0,sau}.

As a shorthand, one can say that the
$k$-order jet manifold $J^kY$ of a bundle $Y\to X$
comprises the equivalence classes
$j^k_xs$, $x\in X$, of sections $s$ of $Y$ identified by the first $k+1$
terms of their Taylor series at a point $x$.  Recall that
a $k$-order differential operator on sections of a bundle $Y$ is defined to be
a bundle morphism of the bundle $J^kY\to X$ to a vector
bundle over $X$.

We restrict ourselves
to the first order  Lagrangian formalism, for
the most of contemporary field models are described by
first order Lagrangian densities. This is not the case for
 the Einstein-Hilbert Lagrangian density of the Einstein's
gravitation theory which belongs to the special class of second order
Lagrangian densities whose Euler-Lagrange equations are however of the order
two as like as in the first order theory. In Part 2 of the work,
we shall discuss particular case of the second order Lagrangian systems in
application to gravitation theory \cite{kru82,nov}.
At the same time, the
Einstein's theory is well-known to be reformulated as the first-order theory in
the Palatini variables when a pseudo-Riemannian metric and a symmetric
connection are considered as independent fields \cite{bor,ped}. Also in
the gauge  theory of gravity, the gravitation interaction is described by two
geometric fields. They are a tetrad gravitational field
and a reduced Lorentz connection which plays the role of a gauge
gravitational potential \cite{heh,sard92}. All gravitational Lagrangian
densities of classical  gravity in these variables are of the first order. In
the absence of fermion fields, one can choose  a general linear connection and
a pseudo-Riemannian metric as a couple of affine-metric gravitational
variables, whereas consideration of the total system of fermion and
gravitational fields calls into play the machinery of composite spinor
bundles \cite{sard92,sard,sard1}.

{}From the mathematical point of view, first order Lagrangian theories are
free from many ambiguities which are present in the higher order ones.

In the first order Lagrangian formalism, the finite-dimensional
configuration space
of fields represented by sections $s$ of a bundle $Y\to X$ is the first order
jet manifold $J^1Y$ of $Y$.  Given fibered coordinates $(x^\m,y^i)$
of $Y$, the jet manifold $J^1Y$ is endowed with the adapted  coordinates
$ (x^\m,y^i,y^i_\m)$. In physical literature, the coordinates $y^i_\m$ are
usually called the velocity coordinates or the derivative coordinates
because of the relation
$$
y^i_\m(j^1_x s) =\dr_\m s^i(x).
$$
The jet manifold $J^1Y$ is endowed with the natural bundle structures
$J^1Y\to Y$ and $J^1Y\to X$. For the sake of convenience, we shall call
$J^1Y\to X$ the configuration bundle and $J^1Y\to Y$ simply the jet
bundle.

A first order
Lagrangian density on $J^1Y$ is defined to be an exterior horizontal density
\be
&& L: J^1Y\to\op\w^nT^*X, \qquad n=\dim X,\\
&&L=\cL(x^\m,y^i,y^i_\m)\om, \qquad \om=dx^1\w ...\w dx^n,
\ee
on the configuration bundle $J^1Y\to X$. It is polinomial in derivative
coordinates $y^i_\m$.

By a differential
conservation law in first order field theories is meant a relation where
the divergence of a current $T$
appears equal to zero, i.e.
\begin{equation}
ds^*T =0 \label{C1}
\end{equation}
where $T$ is a horizontal $(n-1)$-form on the configuration bundle $J^1Y\to X$
and $s$ is a section of the bundle $Y\to X$.

The relation (\ref{C1}) is called a strong conservation law if it
is satisfied identically for all sections $s$ of the bundle $Y\to X$, and it
is termed a weak conservation law if it takes place only on
critical sections, i. e.,  on solutions of field equations. We shall use the
symbol "$\ap$" for weak identities.

It may happen that
the current $T$ conserved weakly is brought into the form
$$
T=W+dU
$$
where $W\ap 0$. In this case,
one says that the current $T$ is reduced to the superpotential $U$
\cite{bor,fat}. For instance, the N\"other currents in gauge theory come to
superpotentials which depend on parameters of gauge transformations that
provide the gauge invariance of N\"oether conservation laws (see Section 7).

In gravitation theory also, conserved currents are reduced to superpotentials
\cite{bor,nov,reg,sza}.  Moreover, we meet the phenomenon of
"hidden energy"  Only the superpotential part of energy-momentum of gravity
and matter is observed when the general covariant transformations are exact
(see Part 2). The Proca field model illustrates that
other parts of energy-momentum of matter display themselves if the
invariance under general covariance transformations is broken, e.g., by a
background world metric (see Section 11).

Usually, one derives the differential conservation
laws from invariance of a Lagrangian density under some group of
symmetries.

Every 1-parameter group $G_t$ of bundle (fiber-to-fiber)
transformations of a bundle $Y\to X$ is
well-known to induce a complete projectable vector field $u$ on $Y$. In
particular, if these are transformations over the identity morphism $\Id_X$
of $X$ (or simply over $X$), the proper vector field $u$ is a vertical vector
field on $Y\to X$ that is tangent to fibers of the bundle $Y$. Conversely,
let $u$ be a projectable vector field on $Y\to X$. Then,
there exists a local 1-parameter group of
transformations of $Y\to X$ which induces the given vector field $u$.

Let $G_t$ be a 1-parameter group of bundle isomorphisms of a bundle $Y\to X$
and $u$ the corresponding vector field on $Y$. One can prove that a
Lagrangian density  $L$ on the configuration space $J^1Y$ is invariant under
these transformations iff its Lie derivative by the lift $\ol u$
of $u$ onto $J^1Y$ is equal to zero:
\begin{equation}
\bL_{\ol u}L=0. \label{C2}
\end{equation}
The equality (\ref{C2}) gives rise to
the weak differential conservation law.
In particular, if $u$ is a vertical vector field on a bundle $Y\to X$,
we get the current conservation law exemplified by
the well-known N\"oether identities in gauge theory.

At the same time, different notions of symmetries
are utilized. One distinguishes usually between
the invariant transformations and the generalized invariant transformations
\cite{bau,kru73,tak}. The latter does not imply the invariance of a Lagrangian
density, but the invariance of the Euler-Lagrange equations (see also
\cite{grig}). For instance, gauge transformations in the familiar Yang-Mills
gauge theory are invariant transformations, whereas in the Chern-Simon gauge
model, they are generalized invariant transformations. There is no
symmetries which would provide the energy-momentum conservation law without
fail. Moreover, in the presense of background fields, e.g., the background
metric field, the Lie derivatives (\ref{C2}) are never equal to zero, and we
have the transformation laws, but not the conservation ones.

Given these circumstances, there are reasons to examine the
Lie derivative of a Lagrangian density by various projectable vector fields
$u$ on the bundle $Y\to X$, without assuming it preliminarly to vanish
\cite{ech1,giach,sard2}. At the same time, every such vector field can be
treated as the generator of a local 1-parameter transformation group.

Let
\[ u=u^\m(x)\dr_\m + u^i(y)\dr_i\]
be a projectable vector field on a bundle $Y\to X$ and $\ol u$ its jet lift
(see (\ref{1.21}) below) onto the configuration space $J^1Y\to X$. Given a
Lagrangian density   $L$, let us compute the Lie derivative ${\bf L}_{\ol
u}L$. We get the canonical decomposition
\begin{equation}
\bL_{\ol u}L= d_HT + u_V\rfloor\cE_L \label{C3}
\end{equation}
where $\cE_L$ is the Euler-Lagrange operator. This is
the well-known first variational formula of the calculus of
variations \cite{kru73,kru86}.
We follow the conventional formulation of the variational problem when
"deformations" of sections $s$ of a bundle $Y\to X$ are induced by
local 1-parameter groups of transformations of the bundle $Y$,
 and the Lie derivatives of Lagrangian densities and their Lepagean
equivalents by the corresponding vector fields on $Y$ are examined
\cite{bau,ded,ech,herm,kru73,kru86,kup}. The
Euler-Lagrange operator  $\cE_L$, by definition, vanishes on the critical
sections of the bundle $Y\to X$, and the equality (\ref{C3}) comes to the weak
identity \begin{equation}
 s^*{\bf L}_{\ol u}L\ap -\frac{d}{dx^\la}[\pi^\la_i(u^\m
\dr_\m s^i-u^i) -u^\la\cL ]\om \label{502}
\end{equation}
where
$$
\pi^\m_i=\dr^\m_i\cL
$$
denotes the Lagrangian momenta. It is the identity (\ref{502}) that most
authors utilize in order to get differential conservation laws in field theory
\cite{ech1,fat,fer,giach,got92,sard2}.

If the Lagrangian density $L$ satisfies the strong equality (\ref{C2}), the
weak identity (\ref{502}) takes the form of the weak conservation law of the
current
$$
T^\la=\pi^\la_i(u^\m
\dr_\m s^i-u^i) -u^\la\cL.
$$
The
gauge theory of exact internal symmetries and the gravitation theory on
bundles of geometric objects are two examples of theories where we can utilize
that the Lie derivatives of Lagrangian densities are equal to zero.

 Note that the current $T$ in the expression
(\ref{C3}) depends on the choice of the Lepagian equivalent of the
Lagrangian density $L$. Different Lepagian equivalents lead to currents
$T$ which differ from each other in the superpotential terms
(see Section 4).

In this work, we analize the different types of transformation and
conservation laws resulting from the weak identity (\ref{502}). The most of
differential transformation and conservation laws of field theory, including
the energy-momentum transformation laws, the general covariance condition and
the above-mentioned N\"oether conservation laws, can be recovered in this
way.

Let $u$ be a vector field on the bundle $Y\to X$ which is projectable onto a
vector field $\tau$ on the base $X$. Then, $u$ can be represented as the sum
of some lift of $\tau$ onto $Y$ and a vertical vector field on $Y$. It
follows that every weak identity (\ref{502}) can be represented as the
superposition of this identity when $u$ is a vertical vector field (i.e., the
identity of the N\"oether type) and that when $u$ is a lift of a vector field
on $X$ anto $Y$.

In general case, a vector field $\tau$ on the
base $X$ gives rise to a vector field on $Y$ only by means of a connection on
the bundle $Y\to X$.  Note that some bundles
admit the canonical lift of vector fields $\tau$ on $X$.

Let
$$\tau=\tau^\la\dr_\la$$ be a vector field on $X$ and
$$
\tau_\G=\tau^\m (\dr_\m+\G^i_\m\dr_i)
$$
its horizontal lift onto $Y$ by a connection $\G$
on $Y\to X$. In this case, the identity (\ref{502}) takes the form
\begin{equation}
 s^*{\bf L}_{\ol\tau_\G}L\ap
-\frac{d}{dx^\la}[\tau^\m \cT_\G{}^\la{}_\m ( s)]\om \label{504}
\end{equation}
where
\[\cT_\G{}^\la{}_\m (s) =\pi^\la_i(\dr_\m s^i -\G^i_\m)
-\dl^\la_\m\cL \]
is the stress-energy-momentum (SEM)
tensor of a field $s$ relative to the connection $\G$.
This is a particular case of
SEM tensors \cite{fer,got92,kij}.

For instance, let us choose the trivial local connection $\G^i_\m=0$.
In this case, the identity (\ref{504}) recovers the well-known conservation law
\[\frac{\dr\cL}{\dr x^\la} +\frac{d}{dx^\la} \cT^\la{}_\m (s)\ap 0\]
of the canonical energy-momentum tensor
$$
\cT^\la{}_\m (s)= \pi^\la_i\dr_\m s^i -\dl^\la_\m\cL.
$$
This however fails to be a
well-behaved mathematical object.

The crucial point lies in the fact that the Lie derivative
$$
{\bf L}_{\ol\tau_\G}L=\{\dr_\m\tau^\m\cL +[\tau^\m\dr_\m
+\tau^\m\G^i_\m\dr_i+(\dr_\la(\tau^\m\G^i_\m)+\tau^\m
y^j_\la\dr_j\G^i_\m -y^i_\m\dr_\la\tau^\m)\dr^\la_i]\cL\}\om
$$
is almost never equal to zero. Therefore, it
is not obvious what SEM tensor is the true energy-momentum tensor, i.e.,
how to choose the connection $\G$ in order to lift vector fields $\tau$ on the
base onto the bundle $Y$.
It may happen that different solutions of the same field equations
require different
SEM tensors in general \cite{sard,sard2}.
At the same time, the SEM tensors relative to different
connections $\G$ and $\G'$ on $Y$ differ from each other in a N\"oether
current associated with the vertical vector field $\tau_\G-\tau_{\G'}$  on the
bundle $Y\to X$.

One finds clear illustration of
this phenomenon in the framework of the time-dependent mechanics
when
$Y$ is a bundle over $\R$
and there is the 1:1
correspondence between the vertical vector fields and the connections on $Y$.
In this case, any first integral of motion appears to be a part of the
stress-energy function relative to a suitable connection \cite{ech}.

It should be mentioned the pecularity of
field theories on the bundles of geometric objects examplified by tensor
bundles and bundles of linear connections. In this case,
there exists the canonical lift $\wt\tau$ of a vector field on the base $X$
onto such a bundle. This lift consists with the horizontal lift of $\tau$ by
means of the connection which meets $\tau$ as the geodesic field.
Calculation
of the Lie derivative  by this lift results in the general
covariance condition of a Lagrangian density. The corresponding SEM
conservation law contains the superpotential term (see Section 8). Being
applied to Einstein's gravitation theory
in its
traditional second-order formulation \cite{kru82,nov} and in the
Palatini variables \cite{bor}, this conservation law leads us to the
well-known Komar superpotential.
In Part 2 of the work, we extend this approach to the affine-metric
gravitation theory and the gauge gravitation theory.
The total gauge model of gravity and fermion fields which calls into
consideration the composite spinor bundle
$$
S\to\Si\to X^4
$$
over the bundle of gravitational fields $\Si\to X^4$. The
pecularity of a Lagrangian system on this composite bundle lies in the fact
that Lagrangian momenta of fermion fields and gravitational fields are
not independent.

Let us emphasize that, in general case of both space-time and internal
symmetries, a Lagrangian density can not satisfy the general covariance
conditions if internal symmetries are broken (see Section 8). As a consequence,
in field theories with broken internal symmetries, no energy-momentum tensor is
conserved. The gauge breaking of the differential energy-momentum
conservation law in the Chern-Simons gauge theory illustrates this phenomenon
(see Section 12).

\section{Geometric preliminaries}

This Section aims to remind the basic notations and operations of the jet
machinery that we shall refer to.

All morphisms throughout are differentiable mappings of
class $C^\infty$. Manifolds are real, Hausdorff,
finite-dimensional, second-countable (hence paracompact), and connected.

We use the conventional symbols
$\otimes$, $\vee$ and $\wedge$ for the tensor, symmetric and
exterior products respectively.
By $\rfloor$ is meant the interior product (contraction) of
multivectors on the right and forms on the left.

The symbols $\dr^A_B$ denote partial derivatives with respect to
coordinates possessing multi-indices $^B_A$.

The tangent bundle $TM$ and the cotangent bundle
$T^*M$ of the manifold $M$ are provided with atlases of the {\it induced
coordinates} $(z^\la, \dot z^\la)$ and $(z^\la,\dot z_\la)$  relative to the
{\it holonomic} fiber bases $\{\dr_\la\}$ and $\{dz^\la\}$ of $TM$ and
$T^*M$ respectively. If $f:M\to M'$ is a manifold mapping, by
\be
&& Tf: TM\to TM', \\
&&\dot z'^\la\circ Tf= \frac{\dr f^\la}{\dr z^\al}\dot z^\al,
\ee
is meant the {\it tangent morphism} to $f$.
\medskip

\noindent
\centerline{\sc Fiber bundles}
\medskip

By a fiber bundle or simply a {\it bundle} is meant a locally trivial fibered
manifold
$$
\pi: Y\to X.
$$
The {\it projection} $\pi$ is a surjective submersion
when both $\pi$ and the tangent morphism $T\pi$ are surjections.
We use the symbols $y$ and $x$ for points of the bundle
$Y$ and its base $X$ respectively.
A bundle $Y\to X$ is endowed with an atlas of
{\it bundle coordinates}
$(x^\la, y^i)$ where $(x^\la)$ are coordinates of the  base $X$. We assume the
manifold $X$ to be oriented.

The differentiable structure on $Y$ possesses some pecularities because of
the fibration $Y\to X$.

The tangent bundle $TY\to Y$ of a bundle $Y$ has the {\it vertical
subbundle}
\[
VY = \Ker T\pi
\]
which consists of the tangent vectors to the fiberes of
$Y$. It is provided
with the induced coordinates $(x^\la,y^i,\dot y^i)$ with respect to
the fiber bases $\{\dr_i\}$.
The {\it vertical cotangent bundle} $V^*Y\to Y$ of $Y$, by definition, is the
vector bundle dual to the vertical tangent bundle $VY\to Y$.
Let us emphasize that this is not a subbundle of the cotangent bundle $T^*Y$ of
$Y$.

For the sake of simplicity, we shall denote
the bundle pullbacks
\[
Y\op\times_X TX, \qquad Y\op\times_X T^*X
\]
simply by $TX$ and $T^*X$.

The vertical tangent bundles $VY$ of many bundles utilized in field
theory are trivial bundles, that is,
$$
VY= Y\op\times_X \overline Y
$$
where $\overline Y\to X$ is a vector bundle. This construction is called the
{\it vertical splitting}.
In particular, a vector bundle $Y\to X$ admits the
canonical vertical splitting
$$
VY=Y\op\times_X Y.
$$
An
affine bundle $Y$ modelled on a vector bundle $\overline Y$ also has the
canonical vertical splitting
$$
VY=Y\op\times_X\overline Y.
$$
\medskip

\noindent
\centerline{\sc Vector fields and differential forms}
\medskip

We
shall deal with the following particular types of vector fields and
differential forms on a bundle $Y\to X$:
\begin{itemize}
\item a {\it projectable vector field}
\[
u=u^\m (x)\dr_\m +u^i(y)\dr_i,
\]
which covers a vector field
\[
\tau_u=u^\mu(x)\dr_\mu
\]
on the base $X$ such that
$$
\tau_u\circ \pi= T\pi\circ u;
$$
\item a {\it vertical vector field}
\[
u= u^i(y)\dr_i: Y\to VY;
\]

\item an {\it exterior horizontal form} $$ \si:Y\to\op\w^r T^*X;$$

\item a {\it tangent-valued horizontal form}
 $$\si:Y\to\op\w^r T^*X\op\otimes_Y TY;$$
\item a vertical-valued {\it soldering form}
\be
&& \si: Y\to T^*X\op\ot_YVY,\\
&& \si=\si_\la^i(y)dx^\la\otimes\dr_i
\ee
and, in particular,
the {\it canonical soldering form}
$$
\theta_X=dx^\la\otimes\dr_\la
$$
on $TX$;

\item the {\it pullback}
$$
\pi^*\si(t_1,\ldots,t_k)=\si(T\pi(t_1),\ldots,T\pi(t_k)), \qquad
t_1,\ldots,t_k\in TY,
$$
of an exterior $k$-form $\si$ on $X$ onto $Y$.
\end{itemize}

Exterior horizontal $n$-forms are called {\it horizontal densities}.
We shall refer to the notation
$$
\om=dx^1\w \cdots\w dx^n, \qquad
 \om_\la =\dr_\la\rfloor\om, \qquad \dr_\m\rfloor\om_\la = \om_{\m\la}.
$$

For any vector field $\tau$ on $X$, we can define its pullback
$$
 \pi^*\tau= \tau\circ\pi: Y\to TX
$$
on $Y$. This is
not a vector field on $Y$, for the tangent bundle $TX$ of $X$ fails to be a
subbundle of the tangent bundle $TY$ of $Y$. One needs a connection on $Y\to
X$ in order to set the imbedding $TX\hookrightarrow TY$.

The {\it Lie derivative} ${\bf L}_u\si$ of an exterior form $\si$
by a vector field $u$ satisfies the relation
\[
{\bf L}_u\si =u\rfloor d\si +du\rfloor\si.
\]
One can refer to this relation as the definition of this Lie
derivative. We recall other useful formulas concerning with Lie derivatives and
differential forms:
\ben
&& \bL_u(\si\w\si') =\bL_u\si\w\si' +\si\w\bL_u\si',\label{C9}\\
&& \bL_u d\si=d(\bL_u\si),\nonumber
\een
and also
\be
&& u\rfloor(\si\w\si') =u\rfloor\si\w\si' +(-1)^{\mid \si\mid}
\si\w u\rfloor\si',\\
&&d\pi^*\si=\pi^*d\si.
\ee

\medskip

\noindent
\centerline{\sc Jet manifolds}
\medskip

The first order {\it jet manifold} $J^1Y$ of a bundle
$Y\to X$ is provided with an atlas of adapted
coordinates $(x^\la,y^i,y_\la^i)$ possessing the transition functions
$$
{y'}^i_\la = (\frac{\dr{y'}^i}{\dr y^j}y_\m^j +
\frac{\dr{y'}^i}{\dr x^\m})\frac{\dr x^\m}{\dr{x'}^\la}.
$$
A glance at this transition functions shows that the
{\it jet bundle}
$$
\pi^1_0:J^1Y\to Y
$$
is an affine bundle. It is modelled on the vector
bundle $$T^*X\op\ot_Y VY\to Y.$$
We shall denote elements of $J^1Y$ by $z$.

There exists the following canonical bundle monomorphisms of the jet bundle
$J^1Y\to Y$:
\begin{itemize}
\item the {\it contact map}
\ben
&&\la_1:J^1Y\op\hookrightarrow_YT^*X \op\ot_Y TY,\nonumber\\
&&\la_1(z)=dx^\la\ot\wh\dr^1_\la=dx^\la\ot(\dr_\la+y^i_\la \dr_i),\label{18}
\een
\item the {\it complementary morphism}
\ben
&&\th_1:J^1Y \hookrightarrow T^*Y\op\otimes_Y VY,\label{24}\\
&&\th_1(z)=\wh{d}y^i \otimes \dr_i=(dy^i- y^i_\la dx^\la)\otimes
\dr_i.\nonumber
\een
\end{itemize}
These canonical morphisms enable us to phrase the jet manifold machinery in the
familiar terms of tangent-valued forms.

The operators
$$
\wh\dr^1_\la=\dr_\la+y^i_\la \dr_i
$$
are usually called  the {\it total derivatives} or the {\it formal
derivatives}. The forms
 $$
\wh{d}y^i=dy^i- y^i_\la dx^\la
$$
are conventionally termed the {\it contact forms}.

We have the so-called jet functor from the category of bundles to the category
of jet manifolds \cite{kol}. It implies the natural prolongation of morphisms
of bundles to morphisms of jet manifolds.

Every bundle morphism of $\Phi: Y\to Y'$
over a diffeomorphism $f$ of $X$ has the {\it jet prolongation} to the bundle
morphism
\[j^1\Phi:J^1Y\to J^1Y',\]
\[ {y'}^i_\m\circ
j^1\Phi=(\dr_\la\Phi^i+\dr_j\Phi^iy^j_\la)\frac{\dr (f^{-1})^\la}{\dr
{x'}^\m},\]
over $\Phi$.

In particular, every section $s$ of a bundle $Y\to X$ admits the
jet prolongation to the section $j^1s$ of the bundle $J^1Y\to X$:
$$
(j^1s)(x)\op =^{\rm def} j_x^1s,
$$
\[
(y^i,y_\la^i)\circ j^1s= (s^i(x),\dr_\la s^i(x)).
\]

Every projectable vector field
\[
u = u^\la (x)\dr_\la + u^i(y)\dr_i
\]
on a bundle $Y\to X$ gives rise to the projectable vector field
\begin{equation}
\ol u =j^1_0u=u^\la\dr_\la + u^i\dr_i + (\dr_\la u^i+y^j_\la\dr_ju^i
- y_\m^i\dr_\la u^\m)\dr_i^\la \label{1.21}
\end{equation}
on the bundle $J^1Y\to X$. If $u$ is a vertical field, its lift
$$
\ol u: J^1Y\to VJ^1Y
$$
consists with the jet prolongation
$$
j^1u: J^1Y\to J^1VY
$$
of $u$ because of the canonical bundle isomorphism
$$
VJ^1Y=J^1VY.
$$

Application of the jet formalism to differential geometry produces
the {\it canonical splitting} of the pullback bundle
\begin{equation}
J^1Y\op\times_Y TY=TX\op\oplus_{J^1Y} VY,\label{C6}
\end{equation}
\[\dot x^\la\dr_\la
+\dot y^i\dr_i =\dot x^\la(\dr_\la +y^i_\la\dr_i) + (\dot y^i-\dot x^\la
y^i_\la)\dr_i,\]
which is induced by the contact map (\ref{18}).

As an immediate consequence of the splitting (\ref{C6}),
one obtains the corresponding splitting
\begin{equation}
\pi^{1*}_0u =u_H +u_V =u^\la (\dr_\la +y^i_\la
\dr_i)+(u^i - u^\la y^i_\la)\dr_i. \label{31}
\end{equation}
of the pullback $\pi^{1*}_0u$ of any vector field
$$
u=u^\la\dr_\la +u^i\dr_i
$$
 on $Y\to X$
onto $J^1Y$. In other words, there is the canonical splitting of
every vector field on a bundle $Y$ over $J^1Y$.

Similarly, the complementary morphism  (\ref{24}) yields
the canonical horizontal splitting
of an exterior 1-form
\begin{equation}
\pi^{1*}_0\si =\si_\la dx^\la + \si^idy^i=(\si_\la +
y^i_\la\si_i)dx^\la+\si_i(dy^i- y^i_\la dx^\la),\label{C22}
\end{equation}
 a tangent-valued horizontal form
\be
&&\phi = dx^{\la_1}\wedge\cdots\wedge dx^{\la_r}\otimes
(\phi_{\la_1\dots\la_r}^\m\dr_\m +
\phi_{\la_1\dots\la_r}^i\dr_i)\\
&&\quad = dx^{\la_1}\wedge\cdots\wedge dx^{\la_r}\otimes
[\phi_{\la_1\dots\la_r}^\m (\dr_\m  +y^i_\m \dr_i)
+(\phi_{\la_1\dots\la_r}^i - \phi_{\la_1\dots\la_r}^\m y^i_\m)\dr_i]
\ee
where we mean summation with respect to ordered collections $\la_1\dots\la_r$
and the canonical soldering 1-form
\begin{equation}
\th_Y=dx^\la\otimes\dr_\la + dy^i\otimes\dr_i
=\la_1 + \th_1=dx^\la\otimes\wh{\dr}^1_\la+\wh d
y^i\otimes\dr_i \label{35}
\end{equation}
on $Y$.
The splitting (\ref{35}) implies the
canonical splitting of the exterior differential
$$
d=d_H+d_V.
$$
 on the pullbacks $\pi^{1*}_0\si$
of horizontal exterior forms
$$
\si=\si_{\la_1\dots\la_r}(y)dx^{\la_1}\wedge\cdots\wedge dx^{\la_r}
$$
 on $Y\to X$ onto $J^1Y$.
In this case, $d_H$ makes the sense of the {\it total differential}
\begin{equation}
 d_H\pi^{1*}_0\si
=\wh\dr^1_\m
\si_{\la_1\dots\la_r}(y)dx^\m\w dx^{\la_1}\wedge\cdots\wedge dx^{\la_r},
\label{C23}
\end{equation}
whereas $d_V$ is the {\it vertical differential}
\begin{equation}
 d_V\pi^{1*}_0\si
 =\dr_i
\phi_{\la_1\dots\la_r}(y)\wh dy^i\w dx^{\la_1}\wedge\cdots\wedge
dx^{\la_r}. \label{C24}
\end{equation}
\medskip

\noindent
\centerline{\sc Connections}
\medskip

Let $\G$ be a section of the jet bundle $J^1Y\to Y$. The replacement
$$y^i_\la=\G^i_\la(y)$$ into the canonical splitting (\ref{C6})
results in the familiar horizontal splitting of the tangent bundle $TY$ of
$Y$ with respect to a connection on $Y$. Moreover,
there is the 1:1 correspondence between the global sections
\[\G =dx^\la\ot(\dr_\la+\G^i_\la\dr_i)\]
of the affine jet bundle $J^1Y\to Y$ and the connections
on the bundle $Y\to X$.
In particular, a linear connection $K$ on the tangent bundle $TX$ of a
manifold $X$ and the dual connection $K^*$ to $K$ on the cotangent bundle
$T^*X$ are given by the coordinate expressions
\be
&& K^\al_\la=-K^\al{}_{\nu\la}(x)\dot x^\nu,\\
&&K^*_{\al\la}=K^\nu{}_{\al\la}(x)\dot x_\nu.
\ee

Connections  on a bundle $Y\to X$ constitute the
affine space modelled on the linear space of soldering 1-forms
on $Y$.
It means that, if
$\G$ is a connection and
$\si$ is a soldering form on a bundle $Y$, its sum
\[
\G+\si=dx^\la\otimes[\dr_\la+(\G^i_\la +\si^i_\la)\dr_i]
\]
is a connection on $Y$. Conversely, if $\G$ and $\G'$ are
connections on a bundle $Y$, then their difference
\[
\G-\G'=(\G^i_\la -{\G'}^i_\la)dx^\la\otimes\dr_i
\]
is a soldering form on $Y$.

Given a connection $\G$, its {\it curvature} is the $VY$-valued
2-form
\be
&&R = \frac12 R^i_{\la\m} dx^\la\w dx^\m\otimes\dr_i,\\
&&R^i_{\la\m}=\dr_\la\G^i_\m -\dr_\m\G^i_\la +\G^j_\la\dr_j\G^i_\m
-\G^j_\m\dr_j\G^i_\la.
\ee

A connection $\G$ on a bundle $Y\to X$ yields the affine
bundle morphism
\be
&&D_\G:J^1Y\ni z\mapsto z-\G(\pi_0^1(z))\in T^*X\op\otimes_Y VY,
\\
&&D_\G =(y^i_\la -\G^i_\la)dx^\la\otimes\dr_i.
\ee
It is  called the {\it covariant differential} relative to the connection $\G$.
The corresponding {\it covariant
derivative} of a section $s$ of $Y$ reads
 $$
\nabla_\G s=D_\G\circ J^1s=[\dr_\la s^i-
(\G\circ s)^i_\la]dx^\la\otimes\dr_i.
$$
In particular, a section $s$ of $Y\to X$
is called the {\it integral section} for a
connection $\G$ on $Y$ if $ \nabla_\G s=0$, that is,
$$
\G\circ s=J^1s.
$$
\bigskip

\noindent
\centerline{\sc Higher order jet manifolds}
\bigskip

Applying the jet functor to the jet bundles, one comes to the higher order
jet manifolds.

  The {\it repeated jet manifold}
$J^1J^1Y$, by definition, is the first order jet manifold of the bundle
$J^1Y\to X$. It is provided with the adapted coordinates
$$(x^\la ,y^i,y^i_\la ,y_{(\m)}^i,y^i_{\la\m}).$$
Its subbundle $ \wh J^2Y$ given by the coordinate relation
$$y^i_{(\la)}= y^i_\la$$ is called the {\it sesquiholonomic jet manifold}.
The {\it second order jet manifold} $J^2Y$ of $Y$, in turn, consists with the
subbundle of $\wh J^2Y$ given by the coordinate relation
$$ y^i_{\la\m}=
y^i_{\m\la}.$$

In particular, the repeated jet prolongation
$j^1j^1s$ of a section $s$ of $Y\to X$ is a  section of the bundle
$J^1J^1Y\to X$. It takes its values into $J^2Y$ and
consists  with the second order jet prolongation $j^2s$ of $s$:
\[
(j^1j^1s)(x)=(j^2s)(x)=j^2_xs.
\]

Given a symmetric connection $K$ on the cotangent bundle $T^*X$, every
connection $\G$ on a bundle $Y\to X$ yields the connection
$$
J\G=dx^\m\otimes
[\dr_\mu+\Gamma^i_\mu\dr_i +(\dr_\la\Gamma^i_\m+
\dr_j\Gamma^i_\mu y^j_\la -
K^\alpha{}_{\la\mu} (y^i_\alpha-\Gamma^i_\alpha))
\dr_i^\la]
$$
on the bundle $J^1Y\to X$.
 Note that the curvature $R$  of a connection $\G$ on a
bundle $Y\to X$ induces the soldering form
$$
\ol\si_R=R^i_{\la\m}dx^\m\otimes\dr^\la_i
$$
on $J^1Y\to X$.

In a similar way, one can describe jet manifolds $J^rY$ of any finite order
$r$ and the {\it infinite order jet pace} $J^\infty Y$ which, by definition, is
the projective limit of the inverse system
$$
X\op\longleftarrow^\pi Y\op\longleftarrow^{\pi^1_0}\ldots J^{r-1}Y
\op\longleftarrow^{\pi^r_{r-1}} J^rY\op\longleftarrow^{\pi^{r+1}_r}\ldots
$$
The infinite order jet space $J^\infty Y$ is endowed with the inverse limit
topology. It can be coordinatized by
$$(x^\al, y^i,\ldots,y^i_{\la_1\ldots\la_r},\ldots)$$
where $\la_1\ldots\la_r$ are collections of numbers modulo rearrangements, but
it fails to be a well-behaved manifold in general. At the same time, one can
introduce the sheaf of smooth functions on $J^\infty Y$ and define the
differential calculus on $J^\infty Y$ \cite{bau,kru73,tak}. Suitable notation
for vector fields, derivatives and differential forms  just as like as in the
finite order case.

 A vector field $u_r$
on the $r$-order jet manifold  $J^rY$ is called projectable if for any $k< r$
there exists a vector field  $u_k$ on $J^kY\to X$ such that
$$
u_k\circ \pi^r_k=T\pi^r_k\circ u_r.
$$
The tangent
morphism $T\pi^r_k$ sends projectable vector fields on $J^rY$ onto the
projectable vector fields on $J^kY$.
In what follows we shall be interested in projectable vector fields $u_r$
which are extension to the higher order jet
manifolds of infinitesimal transformations of the bundle $Y\to X$. The linear
space of projectable vector
fields on $J^\infty Y$, by definition, is the limit of the inverse system of
projectable vector fields on finite order jet manifolds.
As a consequence, every projectable vector field
$$
u=u^\la\dr_\la + u^i\dr_i
$$
on a bundle $Y\to X$
gives rise to a projectable vector field $\ol u^\infty$ on $J^\infty Y$. We
have its canonical decomposition
\ben
&& \ol u^\infty = u^\infty_H + u^\infty_V, \nonumber \\
&& u^\infty_H =u^\la \wh\dr^\infty_\la =u^\la(\dr_\la + y^i_\la\dr_i +...),
\label{C19}\\
&& u^\infty_V = \op\sum_{k=0}^\infty \wh\dr^k_{\la_k}...
\wh\dr^1_{\la_1}u_V{}^i\dr^{\la_1...\la_k}_i, \nonumber
\een
where $u_V$ is the vertical part of the splitting (\ref{31}) of
$\pi^{1*}_0u$. In particular, $u^\infty_H$ is the canonical lift of the vector
field  $\tau= u^\la\dr_\la$ on $X$ onto $J^\infty Y$.

By the same limiting process, the notions of inner product of exterior
forms and projectable vector fields, the Lie bracket of projectable vector
fields and the Lie derivative of exterior forms by projectable vector
fields on $J^\infty Y$ can be introduced. All the usual identities are
satisfied.

In particular, the notion of contact forms is extended to the forms
$$
\wh dy^i_{\la_1\ldots\la_r}=dy^i_{\la_1\ldots\la_r}
-y^i_{\la_1\ldots\la_r\nu}dx^\nu.
$$
Let $\Om^{r,k}$ denote the space of exterior forms  on $J^\infty Y$ which
are of the order $r$ in the horizontal forms $dx^\nu$ and of the order $k$ in
the contact forms. Then, the space $\Om^n$ of exterior $n$-forms on
$J^\infty Y$
admits the unique decomposition
 \begin{equation}
\Om^n=\Om^{n,0}\oplus \Om^{n-1,1}\oplus\ldots\oplus\Om^{0,n}. \label{C7}
\end{equation}
An exterior form is called a {\it k-contact form} if it belongs the space
$\Om^{r,k}$. We denote by $h_k$ the {\it k-contact projection}
$$
h_k: \Om^n\to \Om^{n-k,k}.
$$
For example, the {\it horizontal projection} $h_0$ performs the replacement
$$
dy^i_{\la_1\ldots\la_k}\to y^i_{\la_1\ldots\la_k\nu}dx^\nu.
$$

Note that if $\si$ is an exterior form on the finite order jet manifold
$J^rY$, the decomposition (\ref{C7}) of the pullback
$$
\pi^{\infty *}_r\si =h_0(\si) +\cdots
$$
is reduced to the decomposition
$$
\pi^{r+1 *}_r\si =h_0(\si) +\cdots .
$$
The canonical splitting (\ref{C22}) exemplifies this decomposition in case of
1-forms on $Y$.

Also the exterior differential operator on exterior forms on $J^\infty Y$ is
decomposed into the sum
\begin{equation}
 d=d_H+d_V \label{C25}
\end{equation}
of the total differential operator
$$
 d_H\phi=\wh\dr^\infty_\m
\phi_{\dots}dx^\m\w \dots
$$
and the vertical differential operator
$$
 d_V\phi=\frac{\dr\phi_{\dots}}{\dr y^i_{\la_1\ldots\la_r}} \wh
dy^i_{\la_1\ldots\la_r}\w\dots
$$
These differentials satisfy the cohomology properties
$$
d_Hd_H=0, \qquad d_Vd_V=0, \qquad d_Vd_H +d_Hd_V=0.
$$

Note that if $\si$ is an exterior form on the finite order jet manifold
$J^rY$, the decomposition (\ref{C25}) is reduced to
$$
\pi^{r+1*}_r d\si=d_H\si+d_V\si.
$$
Differentials $d_H$ and $d_V$ are exemplified respectively by the differentials
(\ref{C23}) and (\ref{C24})  in case of $r=0$.

We have the relation
\begin{equation}
h_0(d\si)=d_Hh_0(\si) \label{C32}
\end{equation}

\newpage

\centerline{\large \bf PART 1}
\bigskip

In this Part of the work, we discuss the canonical decomposition (\ref{C3})
which leads to the weak identity (\ref{502}) and analize the different
expressions of this weak identity depending on the different types of vector
fields $u$ on bundles.

\section{Lagrangian formalism of field theory}

Here we do not go deeply into the calculus of variations, but
remind just some basic ingredients in it.

Let $Y\to X$ be a bundle coordinatized by $(x^\la,y^i)$ and $s$ denote its
sections. Let $N$ be an $n$-dimensional compact  submanifold of $X$ with the
boundary $\dr N$. A smooth deformation of a section $s$  with fixed boundary
on a neighborhood $U$ of $N$ is defined to be a one-parameter family $s_t$ of
sections on $U$ such that  $s_0=s$ and $s_t=s$ on some neighborhood of $\dr
N$. This family defines a vertical vector field along the section $s$. There
always exists a vertical field $u$ on $Y$ vanishing on a neighborhood of
$\pi^{-1}(\dr N)$ which induces these transformations. This is called an
admissible vector field.

Let $\rho$ be an exterior $n$-form on the $r$-order jet manifold $J^rY$.
Given a section $s$ of $Y\to X$, let us consider the pullback $j^rs^*\rho$ of
$\rho$ by the $r$-order jet prolongation $j^rs$ of $s$. For the sake of
simplicity, we shall denote the pullbacks $j^rs^*\rho$ by $s^*\rho$.

A section $s$ is called a critical point of the variational problem of the
form $\rho$ iff, for any vertical vector field $u$ on the bundle $Y\to X$
which vanishes over a neighborhood of $\dr N$, we have
\begin{equation}
\op\int_N \, s^*\bL_{\ol u^r}\rho =0 \label{C10}
\end{equation}
where $\ol u^r=j^r_0u$ denotes the $r$-order jet lift of $u$ onto the jet
manifold $J^rY$.

Building on the formula (\ref{C9}),  one can write
$$
s^*\bL_{\ol u^r}\rho= s^*d(\ol u^r\rfloor\rho) + s^*(\ol u^r\rfloor d\rho),
$$
and bring the functional
(\ref{C10}) into the form
\begin{equation}
\op\int_N \, s^*\bL_{\ol u^r}\rho= \op\int_{\dr N}s^*(\ol u^r\rfloor\rho) +
\op\int_N\,s^*(\ol u^r\rfloor d\rho).
\label{C11}
\end{equation}
It may happen however that the relation (\ref{C11}) is not
appropriate in order to get field equations since the second term in the
raght hand side may depend
 on derivatives of components of the vector field $u$ and may contribute
 in the first one integrated over the boundary $\dr N$.

For instance, let $L=\cL\om$
be a first order Lagrangian density on the jet manifold $J^1Y$ of $Y$. We have
\begin{equation}
s^*(\ol u\rfloor dL) =s^*[u^i\dr_i \cL+
\wh\dr_\la u^i\dr_i^\la\cL]\om  \label{C14}
\end{equation}
where $\ol u$ is the jet lift (\ref{1.21}) of $u$ onto $J^1Y$.
The second term in the expression (\ref{C14}) can be rewritten
$$
\wh\dr_\la u^i\dr_i^\la\cL\,\om=
\wh\dr_\la(u^i\dr_i^\la\cL)\,\om - u^i\wh\dr_\la(\dr^\la_i\cL)\om.
$$
It contains the exact form which we should carry to the boundary integral.

This example shows that the condition
$$
 s^*(\ol u^r\rfloor d\rho)=0
$$
is fit on critical sections only if, for any section $s$, the form
$s^*(\ol u^r\rfloor d\rho)$ depends on components of the vector field $u$,
but not of their derivatives. The forms $\rho$ which meet this property, by
definition, are the Lepagian forms.

Indeed, if $\rho$ is an $n$-form on $J^rY$, the following conditions are
equivalent \cite{kru73}.
\begin{itemize}
\item The 1-contact projection $h_1(\pi^{r+1*}_rd\rho)$ is a horizontal form on
the jet bundle $J^{r+1}Y\to Y$, that is,  it is expanded in the terms
$dx^\la$ and $dy^i$ only.
 \item For each projectable vector field $u_r$ on $J^rY$, the
horizontal projection $h_0(u_r\rfloor d\rho)$ depends on the
$\pi^r_0$-projection of $u_r$ only.
\item For every vertical vector field on the jet bundle $J^rY\to Y$, we have
$$
h_0(u_r\rfloor d\rho)=0.
$$
\end{itemize}

Lagrangian densities fail to be the Lepagian forms in general. One however
can replace the variational problem of a Lagrangian density $L$ with the
variational problem of a suitable Lepagian form $\rho_L$ called the Lepagian
equivalent of $L$. This is a Lepagian form on $J^{r+k}Y$ satisfying the
condition
$$
h_0(\rho_L)=\pi^{r+k+1*}_rL.
$$
Hence, we get the equality
$$
\op\int_N\,s^*L=\op\int_N\,s^*\rho_L.
$$
It follows that the variational problem of the Lagrangian density $L$ is
equivalent to the variational problem of its Lepagian equivalent $\r_L$ over
sections $s$ of the bundle $Y\to X$, but not to the variational problem of
$\r_L$ over sections $\ol s$ of $J^{r+k}Y\to X$. Since
$$
\ol s^*h_0(\rho_L)\neq \ol s^*\rho_L,
$$
it may happens that the projection $\pi^{r+k}_r\circ \ol s$ of a critical
section $\ol s$ for the form $\rho_L$ is not a critical section for the
Lagrangian density $L$.

It was proved in several ways that, given an $r$-order Lagrangian density,
its Lepagian equivalents of order $2r-1$ always exist. Their
local expressions are known \cite{got91,kru86}. These Lepagian equivalents of
an $r$-order Lagrangian density constitute an affine space modelled on the
linear space of the Lepagian equivalents of the $r$-order zero Lagrangian
density. The origin of this affine space is usually chosen to be some Cartan
form. It is the $(k\leq 1)$-contact form expanded in horizontal forms and the
contact forms $\wh dy^i_{\mu_1\ldots\mu_{k<r}}$.

At the same time, every Lepagian form $\rho$ is the Lepagian equivalent of
the Lagrangian density
$$
L=h_0(\rho).
$$

Let $\rho_L$ be a Lepagian equivalent of the $r$-order Lagrangian
density $L$. We have the relation
$$
\bL_{\ol u^{r+1}}h_0= h_0\bL_{\ol u^r}
$$
for any projectable vector field $ u$ on the bundle $Y\to X$. In view of this
relation, we find that
\begin{equation}
\pi^{2r*}_r\bL_{\ol u^r}L=h_0(\bL_{\ol u^{2r-1}}\rho_L)=
 h_0(\ol u^{2r-1}\rfloor d\rho_L) + h_0(d\ol u^{2r-1}\rfloor\rho_L)
\label{C21}
\end{equation}
for any projectable vector field $u$ on $Y\to X$.
This relation is called the first variational formula. The
critical sections $s$ of $Y\to X$ for the variational problem of the
$r$-order Lagrangian density $L$, by definition,  must satisfy the relation
\begin{equation}
 s^*\ol u^{2r-1}\rfloor d\rho_L=0 \label{C16}
\end{equation}
for all vertical fields $u$  on $Y\to X$.

At the same time, the critical sections $\ol s$ of the bundle $J^{2r-1}Y\to X$
for the variational problem of the Lepagian equivalent $\rho_L$, by definition,
satisfy the relation
\begin{equation}
 \ol s^*u_r\rfloor d\rho_L=0 \label{C17}
\end{equation}
for all vertical vector fields $u_r$ on $J^{2r-1}Y\to X$. Note that,
given such a critical section, the relation (\ref{C17}) holds also for every
vector field on $J^{2r-1}Y$.

Hereafter, we restrict our consideration to first order Lagrangian densities
$L$ and their Lepagian equivalents on   the first order jet manifold $J^1Y$.
In this case,  the Cartan form $\Xi_L$  consits with
the Poincar\'e-Cartan form and, in comparison with other Lepagian equivalents,
is uniquely defined by the
 Lagrangian density $L$. It is given by the
coordinate expression
\begin{equation}
\Xi_L=\cL\om +\pi^\la_i(dy^i -y^i_\mu dx^\mu)\w\om_\la. \label{303}
\end{equation}
We choose it as the origin of the affine space of the Lepagian equivalents of
the first order Lagrangian density $L$.
Then, the general local expression for
a Lepagian equivalent of the first order Lagrangian density $L$ reads
\begin{equation}
\r_L=\Xi_L - (\wh\dr_\nu c^{\mu\nu}_i\wh dy^i + c^{\mu\nu}_i\wh dy^i_\nu)
\w\om_\mu + \chi  \label{C56}
\end{equation}
where $c^{\mu\nu}_i=-c^{\nu\mu}_i$ are skew-symmetric functions on $Y$
and $\chi$ is a $(k>1)$-contact form on $J^1Y$ \cite{got91}.

Let us consider the variational problem of the  Poincar\'e-Cartan form
$\Xi_L$ (\ref{303}) over section $\ol s$ of the configuration bundle $J^1Y\to
X$. The critical sections $\ol s$, by definition, obey the relation
\begin{equation}
 \ol s^*u_1\rfloor d\Xi_L=0 \label{C28}
\end{equation}
for all vertical vector fields
$$
u_1=u^i\dr_i +u^i_\mu\dr^\mu_i
$$
on $J^1Y\to X$. Due to arbitrariness of the functions $u^i$ and
$u^i_\mu$, the critical section $\ol s$ ought to satisfy the system of first
order differential equations
\bea
&&\dr^\mu_j\pi^\la_i (\dr_\la\ol s^i - \ol s^i_\la) =0, \label{S8a}\\
&& \dr_i\cL-(\dr_\la+\ol s^j_\la\dr_j
+\dr_\la\ol s^j_\m\dr^\m_j)\dr^\la_i\cL
+\dr_i\pi^\la_j(\dr_\la\ol s^j -\ol s^j_\la)=0. \label{S8b}
\eea
They are called the Cartan equations.
Note that, given a solution $\ol s$ of the Cartan equations, the relation
(\ref{C28}) holds for all vector fields on $J^1Y$.

Now turn to the variational problem of a first order Lagrangian density $L$.
In this case, the first variational formula (\ref{C21}) reads
\begin{equation}
\pi^{2*}_1\bL_{\ol u}L=
 h_0( \ol u\rfloor d\rho_L) + h_0(d \ol u\rfloor\rho_L)\label{C26}
\end{equation}
for any projectable vector field $u$ on $Y$.
After simple calculations, we get
\begin{equation}
\pi^{2*}_1\bL_{\ol u}L=
 u_V\rfloor \cE_L + h_0(d \ol u\rfloor\rho_L) \label{C30}
\end{equation}
where
$$
u_V=(u\rfloor\wh dy^i)\dr_i
$$
is the vertical part of the canonical splitting (\ref{31})
of the pullback $\pi^{1*}_0u$ of $u$ onto
$J^1Y$, and
\begin{equation}
\cE_L=
 [\dr_i-(\dr_\la +y^j_\la\dr_j+y^j_{\m\la}\dr^\m_j)
\dr^\la_i]\cL dy^i\w\om=\dl_i\cL dy^i\w\om \label{305}
\end{equation}
is the Euler-Lagrange operator associated with the Lagrangian density $L$.
Its coefficients
$\dl_i\cL$ are called the variational derivatives of the
Lagrangian density $L$.
It
follows that critical sections $s$ of the bundle $Y\to X$ for the variational
problem of the first order Lagrangian density $L$ satisfy
the second order Euler-Lagrange equations
\begin{equation}
\dr_i\cL-(\dr_\la+\dr_\la s^j\dr_j
+\dr_\la\dr_\m s^j \dr^\m_j)\dr^\la_i\cL=0.\label{2.29}
\end{equation}

It is readily observed that the Euler-Lagrange equations (\ref{2.29}) are
equivalent to the Cartan equations (\ref{S8a}) and (\ref{S8b}) when the
Lagrangian density is regular. In this case, the Cartan equations admit only
holonomic solutions $\ol s=j^1s$. It follows that the variational problem of
a Lagrangian density $L$ is equivalent to the variational problem of its
Poincar\'e-Cartan equivalent when $L$ is regular, otherwise in general case.

Note that different Lagrangian densities  $L$ and $L'$ can result in the
same Euler-Lagrange operator $\cE_L=\cE_{L'}$. It means that they differ
from each other in the Lagrangian density $L_0$ whose
Euler-Lagrange operator $\cE_{L_0}$
is equal to zero.
It can be proved \cite{kru73} that the nesessary and
suffisient condition for $\cE_{L_0}=0$ is that there exists a closed $n$-form
$\e$ on the bundle $Y$ such that
\begin{equation}
L_0=h_0(\e).\label{C44}
\end{equation}
Note that this form $\e$ is not the
Poincar\'e-Cartan form in general.

Any closed form $\e$ on the jet
manifold $J^1Y$ is a Lepagean form. Let $L$ be a Lagrangian density $L$ and
$\rho_L$ its Lepagian equivalent. Then, the Lepagian form $\rho_L+\e$ is the
Lepagean equivalent of the Lagrangian density
$$
L'=L+h_0(\e)
$$
which, as like as the  Lagrangian density $L$, leads to the same
Euler-Lagrange operator. In particular, $L'=L$ if $\e$ is a contact
form.

Remark that, since $J^1Y\to Y$ is an affine bundle, the de Rham cohomologies
of $J^1Y$ consists with those of $Y$ \cite{bau}. It means that every closed
exterior form on $J^1Y$ is the sum of the pullback of a closed exterior form
on $Y$ onto $J^1Y$ and an exact form on $J^1Y$.

\section{Conservation laws}

In the first order Lagrangian theory, we have the following differential
transformation and conservation laws on solutions of the  Euler-Lagrange
equations (\ref{2.29}).

Let $L$ be a Lagrangian density on the jet manifold $J^1Y$. For the sake of
simplicity, we shall denote the pullback $\pi^{1*}_0L$ of $L$ onto $J^2Y$ by
the same symbol $L$.

Let $u$ be a projectable vector field on $Y\to X$ and $\ol u$ its jet lift
(\ref{1.21}) onto the configuration bundle $J^1Y\to X$. We recall that the
vector field $u$ is associated with some 1-parameter group of transformations
of the bundle $Y$.

Let us compute the Lie derivative ${\bf L}_{\ol u}L$ of the horizontal
density $L$ when its Lepagian equivalent is chosen to be the Poincar\'e-Cartan
form $\Xi_L$ (\ref{303}). We recover the first variational formula
(\ref{C30}) in case of projectable vector fields on $Y$ as follows
\cite{giach,sard2}:
\begin{equation}
\bL_{\ol u}L=u_V\rfloor\cE_L + h_0(d\ol u\rfloor\Xi_L). \label{C31}
\end{equation}

Since the  Poincar\'e-Cartan form $\Xi_L$ is a horizontal form on the jet
bundle $J^1Y\to Y$ and the relation (\ref{C32}) holds, the formula
(\ref{C31}) takes the form
\begin{equation}
\bL_{\ol u}L= u_V\rfloor\cE_L + d_H h_0(u\rfloor\Xi_L). \label{C45}
\end{equation}
Being restricted to the kernel
$$
 [\dr_i-(\dr_\la +y^j_\la\dr_j+y^j_{\m\la}\dr^\m_j)
\dr^\la_i]\cL =0
$$
of the Euler-Lagrange operator $\cE_L$ (\ref{305}), the equality (\ref{C45})
reduces to the weak identity
\begin{equation}
\bL_{\ol u}L\ap d_Hh_0( u\rfloor\Xi_L), \label{J4}
\end{equation}
$$
\dr_\la u^\la\cL +[u^\la\dr_\la+
u^i\dr_i +(\dr_\la u^i +y^j_\la\dr_ju^i -y^i_\m\dr_\la u^\m)\dr^\la_i]\cL
\ap \wh \dr_\la[\pi^\la_i(u^i-u^\m y^i_\m )+u^\la\cL],
$$
$$
\wh\dr_\la =\dr_\la +y^i_\la\dr_i+y^i_{\m\la}\dr^\m_i.
$$

On solutions $s$ of the Euler-Lagrange equations, the
weak identity (\ref{J4}) comes to the weak differential transformation law
\begin{equation}
s^*{\bf L}_{\ol u}L\ap d(s^*u\rfloor\Xi_L)\label{C40}
\end{equation}
which takes the coordinate form (\ref{502}).

Note that, in order to obtain the differential transformation laws on
solutions of a given system of Euler-Lagrange equations (\ref{2.29}), one can
examine other Lepagian equivalents $\rho_L$ of the Lagrangian density $L$,
besides the Poincar\'e-Cartan form $\Xi_L$. In this case, the first
variational formula (\ref{C26}) and the corresponding weak identity
$$
\bL_{\ol u}L\ap h_0(d\ol u\rfloor\r_L)
$$
differ from relations (\ref{C31}) and (\ref{J4}) respectively in the strong
identity
\begin{equation}
0 = h_0(d\ol u\rfloor\ve)=d_Hh_0(\ol u\rfloor\ve)\label{C46}
\end{equation}
where $\r_L=\Xi_L +\ve$. From the physical point of view, it means that
different Lepagian equivalents result in different superpotentials
$h_0(\ol u\rfloor\ve)$
in the transformation laws.

In virtue of the formula (\ref{C56}), the form
$\ve$ in the identity (\ref{C46}) has the coordinate expression
$$
\ve =- (\wh\dr_\nu c^{\mu\nu}_i\wh dy^i + c^{\mu\nu}_i\wh dy^i_\nu)
\w\om_\mu + \chi.
$$
It is the general local expression for Lepagian equivalents of the zero
Lagrangian density. We have
$$
h_0(\ol u\rfloor\ve)=\wh\dr_\nu[(u^i-y^i_\la u^\la)c_i^{\m\nu}]\om_\m.
$$

 One can
consider also other Lagrangian densities $L'$ which possess the same
Euler-Lagrange operator $\cE_L$.  Then in virtue of the relation
(\ref{C44}),
the first
variational formula and the corresponding weak identity
differ from relations (\ref{C31}) and (\ref{J4}) respectively in the strong
identity
\begin{equation}
\bL_{\ol u}h_0(\e) = h_0(d\ol u\rfloor\e)\label{C47}
\end{equation}
where $\e$ is some closed exterior form on $Y$. However, if the form
$h_0(\e)$ possesses the same symmetries as the Lagrangian density $L$ only,
the contribution of the strong identity (\ref{C47}) into the weak identity
(\ref{J4}) is not tautological.

It is readily observed that the weak identity (\ref{J4}) is linear in
the vector field $u$, and we can consider superposition of different weak
identities (\ref{J4}) corresponding to different vector fields $u$. For
instance, if $u$ and $u'$ are projectable vector fields on the bundle $Y\to X$
which are projected onto the same vector field on the base $X$, their
difference $u-u'$ is a vertical vector field on $Y\to X$. Accordingly, the
difference of the weak identity (\ref{J4}) with respect the vector fields $u$
and $u'$ results in the weak identity (\ref{J4}) with
respect to the vertical vector field $u-u'$.

Now let us consider the case when a Lagrangian density $L$ depends on
background fields. We define such a Lagrangian density as the pullback of the
Lagrangian density $L_{\rm tot}$ on the total configuration space by some fixed
sections $\f(x)$
describing background fields.

Let us consider the product
\begin{equation}
Y_{\rm tot}=Y\op\times_X Y'\label{C41}
\end{equation}
of the bundle $Y$ whose sections are dynamic fields and the bundle $Y'$ whose
sections $\f$ play the role of background fields.
Let the bundles $Y$ and $Y'$ be coordinatized by $(x^\la, y^i)$ and
$(x^\la, y^A)$ respectively. The Lagrangian density $L_{\rm tot}$ is defined
on the total configuration space $J^1Y_{\rm tot}$.

Let $u$ be a projectable vector field on $Y_{\rm tot}$ which is also
projectable with respect to projection
$$
Y\op\times_X Y'\to Y'.
$$
It has the coordinate form
$$
u=u^\la(x)\dr_\la + u^A(x^\mu,y^B)\dr_A + u^i(x^\mu,y^B, y^j)\dr_i.
$$
It is the natural requirement which means that
transformations of background fields are independent on dynamic fields.

Calculating  the  Lie derivative of the Lagrangian density $L_{\rm tot}$ by
this vector field, we get the equality
\be
&&\dr_\la u^\la\cL_{\rm tot} +[u^\la\dr_\la+  u^A\dr_A +
u^i\dr_i +(\dr_\la u^A +y^B_\la\dr_Bu^A -y^A_\m\dr_\la
u^\m)\dr^\la_A +\\
&& \qquad (\dr_\la u^i  +y^B_\la\dr_Bu^i +y^j_\la\dr_ju^i -y^i_\m\dr_\la
u^\m)\dr^\la_i]\cL_{\rm tot} = \wh \dr_\la[\pi^\la_i(u^i-u^\m y^i_\m )
+u^\la\cL_{\rm tot}]+\\
&&\qquad (u^i-y^i_\la u^\la)(\dr_i -\wh\dr_\la\dr^\la_i)\cL_{\rm tot} +
(u^A-y^A_\la u^\la)\dr_A\cL_{\rm tot} + \pi^\la_A\wh\dr_\la (u^A-y^A_\mu
u^\mu). \ee
It is readily observed that this equality is brought into the form
\be
&&\dr_\la u^\la\cL_{\rm tot} +[u^\la(\dr_\la+ y^B_\la\dr_B
+y^B_{\m\la}\dr_B^\m)\cL + u^i\dr_i +(\wh\dr_\la u^i
 -y^i_\m\dr_\la u^\m)\dr^\la_i]\cL_{\rm tot} = \\
&& \qquad \wh\dr_\la[\pi^\la_i(u^i-u^\m y^i_\m )+u^\la\cL_{\rm tot}] +
(u^i-y^i_\la u^\la)(\dr_i -\wh\dr_\la\dr^\la_i)\cL_{\rm tot}.
\ee
The pullback of this equality to the bundle $Y\to X$ by sections $\f^A(x)$ of
the  bundle $Y'$ which describe the background fields results in the familiar
expression (\ref{C45}) and the familiar weak identity (\ref{J4}) for the
Lagrangian density
$$
L=\f^*L_{\rm tot};
$$
now the partial derivative $\dr_\la$ can be written as
$$
\dr_\la= \wt\dr_\la + \dr_\la \f^B\dr_B+ \dr_\la\dr_\m\f^B\dr_B^\m
$$
where $\wt\dr_\la$ denote the partial derivatives with respect to the
coordinates $x^\la$ on which the Lagrangian density $L_{\rm tot}$ depends in
the explicit form.

Note that Lagrangian densities of field models almost never depend explicitly
on the world coordinates $x^\la$. At the same time, almost all field models
describe fields in the presence of a background world metric $g$ on the base
$X$, except topological field theories whose classical Lagrangian densities are
independent on $g$ \cite{bir} and the gravitation theory where a world
metric $g$ is a dynamical field.

By a world metric on $X$ is meant a nondegenerate fiber metric $g^{\m\n}$ in
cotangent and tangent bundles of $X$. In this case, the partial derivative
$\dr_\la\cL$ in the weak identity (\ref{J4}) contains the term $$
\frac{\dr\cL}{\dr g^{\m\n}}\dr_\la g^{\m\n},
$$
so that the metric energy-momentum tensor of fields
$$
t_{\m\n}\sqrt{\mid g\mid}=2\frac{\dr\cL}{\dr g^{\m\n}}, \qquad
\mid g\mid=\mid\det(g_{\m\n})\mid.
 $$
is called into play.

The weak identity (\ref{J4}) and the weak transformation law (\ref{C40}) are
basic for our analysis of differential transformation and
conservation laws in field theory.

In particular, one says that an isomorphism $\Phi$ of the bundle $Y\to X$ is an
invariant transformation if its jet prolongation $j^1\Phi$ preserves the
Lagrangian density $L$, that is,
$$
j^{1*}\Phi L =L.
$$
Let $u$ be a projectable vector field on $Y\to X$. The corresponding local
1-parameter groups of isomorphisms of $Y$ are
invariant transformations iff the strong equality
$$
\bL_{\ol u}L= 0
$$
holds.
In this case, we have the corresponding weak conservation law
\begin{equation}
 d(s^*u\rfloor\Xi_L)\ap 0.\label{C42}
\end{equation}

An isomorphism $\Phi$ of the bundle $Y\to X$ is called the generalized
invariant transformation if it preserves the
Euler-Lagrange operator $\cE_L$.
Let $u$ be a projectable vector field on $Y\to X$. The corresponding local
 isomorphisms of $Y$ are
generalized invariant transformations iff
$$
\bL_{\ol u}L= h_0(\e)
$$
where $\e$ is a closed $n$-form on the bundle $Y\to X$.
In this case, the weak transformation law (\ref{C40}) reads
$$
s^*\e\ap d(s^*u\rfloor\Xi_L)
$$
for every critical section $s$ of $Y\to X$.
In particular, if $\e=d\ve$ is an exact form, we get the weak conservation law
$$
d(s^*(u\rfloor\Xi_L-\ve)) \ap 0.
$$

In particular, gauge transformations in gauge theory on a 3-dimensional base
$X$ are the invariant transformations if $L$ is the Yang-Mills Lagrangian
density and they are the generalized invariant transformations if $L$ is the
Chern-Simons one (see Section 12).

\section{SEM conservation laws}

Every projectable vector field $u$ on the bundle $Y\to X$ which covers a
vector field $\tau$ on the base $X$ is represented as the sum of a vertical
vector field on $Y\to X$ and some lift of $\tau$ onto $Y$. Hence, any
differential transformation law (\ref{C40}) can be represented as a
superposition of some transformation law associated with a vertical
vector field on the bundle $Y\to X$ and the one induced by the lift of a vector
field on the base $X$ onto $Y$. Therefore, we can reduce our consideration to
transformation laws associated with these two types of vector fields on $Y$.

Vertical vector fields result in transformation and conservation laws of
N\"oether currents. Section 6 is devoted to them.

In general case, a vector field $\tau$ on a base
$X$ gives rise to a vector field on $Y$ only by means of some connection on
the bundle $Y\to X$. Such lifts result in the transformation laws of the SEM
tensors.

Given a bundle $Y\to X$, let $\tau$ be a vector field on $X$ and
$$
\tau_\G=\tau\rfloor\G=\tau^\m (\dr_\m+\G^i_\m\dr_i)
$$
its horizontal lift onto $Y\to X$ by means of a
connection
$$
\G=dx^\m\ot (\dr_\m+\G^i_\m\dr_i)
$$
on $Y$.
In this case, the weak identity (\ref{J4}) is written
\ben
&&\dr_\m\tau^\m\cL +
[\tau^\m\dr_\m
+\tau^\m\G^i_\m\dr_i +(\dr_\la(\tau^\m\G^i_\m)
+\tau^\m y^j_\la\dr_j\G^i_\m-y^i_\m\dr_\la\tau^\m)\dr^\la_i]\cL - \nonumber\\
&& \qquad \wh\dr_\la
[\pi^\la_i(\tau^\m \G^i_\m-\tau^\m y^i_\m)+\dl^\la_\m\tau^\m\cL]\ap
0.\label{C58}
\een
One can simplify it as follows:
$$
\tau^\m\{[\dr_\m
+\G^i_\m\dr_i +(\dr_\la\G^i_\m +y^j_\la\dr_j\G^i_\m)\dr^\la_i]\cL-
\wh\dr_\la [\pi^\la_i(\G^i_\m-y^i_\m)+\dl^\la_\m\cL]\}\ap 0.
$$
Let us emphasize that this relation takes place for arbitrary  vector field
$\tau$ on $X$. Therefore, it is equivalent to the
 system of the weak identities
\begin{equation}
[\dr_\m
+\G^i_\m\dr_i +(\dr_\la\G^i_\m +y^j_\la\dr_j\G^i_\m)\dr^\la_i]\cL-
\wh\dr_\la [\pi^\la_i(\G^i_\m-y^i_\m)+\dl^\la_\m\cL]\ap 0. \label{C52}
\end{equation}

On solutions $s$ of the Euler-Lagrange equations, the weak
identity (\ref{C58}) comes to the weak transformation law
$$
 s^*{\bf L}_{\ol\tau_\G}L
+\frac{d}{dx^\la}[\tau^\m \cT_\G{}^\la{}_\m ( s)]\om \ap 0
$$
and to the equivalent system of the weak transformation laws
 \begin{equation}
[\dr_\m
+\G^i_\m\dr_i +(\dr_\la\G^i_\m +\dr_\la s^j\dr_j\G^i_\m)\dr^\la_i]\cL+
\frac{d}{dx^\la} [\pi^\la_i(\dr_\m s^i-\G^i_\m)-\dl^\la_\m\cL]\ap 0
\label{C53}
\end{equation}
 where $\cT_\G{}^\la{}_\m (s)$ is the SEM tensor given by the
 components  of the $T^*X$-valued $(n-1)$-form
$$
\cT_\G(s)=-(\G\rfloor\Xi_L)\circ s =[\pi^\la_i(\dr_\m s^i-\G^i_\m)
-\dl^\la_\m\cL]dx^\m\ot\om_\la
$$
on $X$.

It is readily observed that the first and the second terms in (\ref{C53})
taken separately fail to be well-behaved objects. Therefore, only their
combination may result in the satisfactory transformation or conservation
law.

For instance, let a Lagrangian density $L$ depend on a background
metric $g$ on the base $X$. In this case, we have
$$
\dr_\m\cL= -t_\bt^\al\sqrt{\mid g\mid}\{^\bt{}_{\m\al}\}
$$
where $\{^\bt{}_{\m\al}\}$ are the Christoffel symbols of the metric $g$ and
$$
t_\bt^\al=g^{\al\g}t_{\g\bt},
$$
by definition,
is the metric energy-momentum tensor. Then, the weak transformation law
(\ref{C53}) takes the form
$$
-t_\bt^\al\sqrt{\mid g\mid}\{^\bt{}_{\m\al}\}+
[\G^i_\m\dr_i +(\dr_\la\G^i_\m +\dr_\la s^j\dr_j\G^i_\m)\dr^\la_i]\cL+
\frac{d}{dx^\la} [\pi^\la_i(\dr_\m s^i -\G^i_\m)-\dl^\la_\m\cL]\ap 0,
$$
and, under suitable conditions of symmetries of the Lagrangian density $L$,
it may come to the covariant conservation law
$$
\nabla_\al t^\al_\bt =0
$$
where $\nabla_\al$ denotes the covariant derivative relative to the connection
$\{^\bt{}_{\m\al}\}$.

Note that, if we consider another Lepagian equivalent (\ref{C56}) of the
Lagrangian density $L$, the SEM transformation law takes the form
$$
 s^*{\bf L}_{\ol\tau_\G}L
+\frac{d}{dx^\la}[\tau^\m {\cT'}_\G{}^\la{}_\m ( s)]\om \ap 0
$$
where
$$
{\cT'}_\G{}^\la{}_\m =\cT_\G{}^\la{}_\m -\frac{d}{dx^\nu}[( \dr_\m
s^i-\G^i_\m)c_i^{\la\nu}],
$$
that is,  the SEM tensors ${\cT'}_\G{}^\la{}_\m$  and
$\cT_\G{}^\la{}_\m$  differ from each other in the superpotential-type term
$$
-\frac{d}{dx^\nu}[( \dr_\m
s^i-\G^i_\m)c_i^{\la\nu}].
$$

In particular, if the bundle $Y$ is provided with a fiber metric $a^Y_{ij}$,
one can choose
$$
c^{\m\nu}_i=a^Y_{ij}g^{\m\al}g^{\nu\bt}R^j_{\al\bt}
$$
where $R$ is the curvature of the connection $\G$ on the bundle $Y$ and $g$
is a metric on $X$. In this case, the superpotential contribution into the
SEM tensor is
$$
-\frac{d}{dx^\nu}[a^Y_{ij}g^{\la\al}g^{\nu\bt}(\dr_\m
s^i-\G^i_\m)R^j_{\al\bt}].
$$

Let us now consider the weak identity (\ref{C58})
 when a vector field $\tau$ on the base $X$ gives rise to a vector field on $Y$
by means of different connections $\G$ and $\G'$ on $Y\to X$. Their difference
result in the weak identity
\begin{equation}
[\tau^\m\si^i_\m\dr_i +(\dr_\la(\tau^\m\si^i_\m) +y^j_\la\dr_j(\tau^\m
\si^i_\m))\dr^\la_i]\cL - \wh\dr_\la [\pi^\la_i\tau^\m\si^i_\m]\ap 0
\label{C57}
\end{equation}
where $\si =\G'-\G$ is a soldering form on the bundle $Y\to X$ and
\begin{equation}
\tau\rfloor\si =\tau^\m\si^i_\m\dr_i \label{C59}
\end{equation}
is a vertical vector field. It is readily observed that the identity
(\ref{C57}) is exactly the weak identity (\ref{J4}) in case of the vertival
vector field (\ref{C59}).

It follows that every SEM transformation law contains a
N\"oether transformation law. Conversely, every N\"oether transformation
law associated with a vertical vector field $u_V$ on $Y\to X$ can be obtained
as the difference of two SEM transformation laws if the vector field $u_V$
takes the form
$$
u_V=\tau\rfloor\si
$$
where $\si$ is some soldering form on $Y$ and $\tau$ is a vector field on
$X$. In field theory, this representation fails to be unique.
On the contrary, in Newtonian mechanics there is the 1:1 correspondence
between the vertical vector fields and the soldering forms on the bundle
$$
\R\times F\to F.
$$

 Note tha one can consider the pullback of the first order Lagrangian density
$L$ and their Lepagian equivalents onto the infinite order jet space $J^\infty
Y$. In this case, there exists the canonical lift $\tau^\infty_H$ (\ref{C19})
of a vector field $\tau$ on $X$ onto $J^\infty Y$. One can treat this lift as
the horizontal lift of $\tau$ by means of the canonical connection
$$
\G_\infty =dx^\m\ot(\dr_\m +y^i\dr_i +y^i_\la\dr^\la_i +\cdots)
$$
on the bundle $J^\infty Y\to X$.

\section{Stress-energy conservation laws in mechanics}

Let us consider a bundle $Y\to \R$. The first order jet manifold $J^1Y$ of
$Y$ plays the role of the configuration space of the Newtonian mechanics
\cite{giach3}.

For the sake of simplicity, we choose a connection $\G_0$ on the bundle
$Y$ and set the corresponding splitting
 \begin{equation}
Y=\R\times F \label{C64}
\end{equation}
coordinatized by $(t,y^i)$. Then we have
\begin{equation}
J^1Y=\R\times ATF \label{C60}
\end{equation}
where $ATF$ denotes the affine tangent bundle of the manifold $F$.
This is the affine bundle modelled on the vector bundle
$$
\R\times TF\to \R\times F.
$$
It is coordinatized by $(t,y^i, y^i_t)$ where $y^i_t$ are the
affine fiber coordinates of $ATF$ associated with the induced fiber coordinates
$\dot y^i$ of the tangent bundle $TF$.
 The derivative coordinates
$y^i_t$ make the sense of velocities with respect to the reference frame
defined by the connection $\G_0$ and the splitting (\ref{C64}).
Note that, in mechanics, Lagrangian densities are
polynomial in the affine coordinates $y^i_t$. Therefore, they factorize in
the following way
 $$
L: \R\times ATF\op\to^{D_\G} \R\times TF\to \R
$$
where  $D_\G$ is the covariant differential relative to some connection $\G$
on $Y$. From the physical point of view, such a connection $\G$ defines some
reference frame so that the quantities
$$
\dot y^i\circ D_\G=y^i_t-\G^i
$$
can be treated as velocities with respect to this refernce frame. In
particular,
$$
\dot y^i\circ D_{\G_0}=y^i_t.
$$

Every connection $\G$ on the bundle $Y$ (\ref{C64}) takes the form
\ben
&& \G: \R\times F\op\to_\R \R\times ATF,\nonumber\\
&& \G=dt\ot (\dr_t +\G^i\dr_i)=\G_0 +\si,\label{C70}
\een
where
$$
\si:\R\times F\to V(\R\times F)=\R\times TF
$$
 is a soldering form. It is readily observed that there exists the 1:1
correspondence
\begin{equation}
u=\dr_t\rfloor\si =\si^i\dr_i \label{C71}
\end{equation}
between the soldering forms $\si$ and the
vertical vector fields $u$ on the bundle (\ref{C64}) or, that is the same,
between the soldering forms $\si$ on the bundle (\ref{C64}) and the
time-dependent vector fields
$$
{\rm pr}_2\circ\si:\R\times F\to TF
$$
on the manifold $F$.

Let us consider the canonical vector field $\tau=\dr_t$ on the base $\R$ of the
bundle (\ref{C64}). Its horizontal lift $\tau_\G$ onto $Y$ by means of the
connection $\G$ (\ref{C70}) reads
$$
\tau_\G=\dr_t +\G^i\dr_i = \tau_0 + u
$$
where $\tau_0$ denotes the horizontal lift of $\tau$ by means of the
trivial connection $\G_0$ and $u$ is the vertical vector field (\ref{C71})
on $Y$. The jet lift (\ref{1.21}) of $\tau_\G$ onto the jet manifold
(\ref{C60}) is written
\begin{equation}
\ol \tau_\G= \dr_t +\G^i\dr_i +(\dr_t\G^i +y^j_t\dr_j\G^i)\dr_i^t. \label{C65}
\end{equation}

Let $L=\cL dt$ be a Lagrangian density on the configuration space (\ref{C60}).
Remind that, since $n=1$, there exists the unique Lepagian equivalent
$$
\Xi_L=(L-\pi_iy^i_t)dt +\pi_idy^i
$$
of $L$.

Computing the Lie derivative of the Lagrangian density $L$ by the vector field
(\ref{C65}), we get the following  weak transformation law (\ref{C53}) on
solutions $s^i(t)$ of the Lagrange equation:
\begin{equation}
[\dr_t +\G^i\dr_i +(\dr_t\G^i +\dr_t s^j\dr_j\G^i)\dr_i^t]\cL +\frac{d}{dt}
[\pi_i(\dr_t s^i -\G^i) -\cL]\ap 0. \label{C66}
\end{equation}
We call it the stress-energy (SE) transformation law and
\begin{equation}
E_\G =\pi_i(\dr_t s^i -\G^i) -\cL \label{C67}
\end{equation}
the SE function relative to the connection $\G$.

If $\G=\G_0$, the
transformation law (\ref{C66}) takes the familiar form of the energy
transformation law in the Lagrangian mechanics
\begin{equation}
\dr_t\cL+
\frac{d}{dt}(\pi_i\dot s^i-\cL)\ap 0. \label{J1}
\end{equation}

Let us consider the difference of the transformation laws (\ref{C66}) and
(\ref{J1}):
$$
[\si^i\dr_i +(\dr_t\si^i +\dr_t s^j\dr_j\si^i)\dr_i^t]\cL -\frac{d}{dt}
(\pi_i\si^i)\ap 0.
$$
It is exactly the weak transformation law (\ref{C40}) of the
N\"oether current
$$
J=\pi_i\si^i
$$
corresponding the vertical vector field (\ref{C71}).

Moreover, every N\"oether transformation law in mechanics can be recovered in
this way due to the 1:1 correspondence (\ref{C71})  between vertical
vector fields and soldering forms on the bundle (\ref{C64}). It follows
that, in Lagrangian mechanics, every first integral of motion is a part of the
SE function relative to the  suitable connection (\ref{C70}).

The following example may help to clarify the physical meaning of the SE
function $E_\G$ (\ref{C67}) with respect to the connection $\G$.

Let us consider the connection
\begin{equation}
\G^i= a^it \label{C90}
\end{equation}
on the bundle (\ref{C64}) which defines the accelerated reference frame
 with respect to $\G_0$. Consider  the Lagrangian density
$$
L = \frac12M(y^i_t-a^it)^2dt
$$
which describes the free massive particle relative to the reference frame
$\G$. It is easy to see that  the SE function
relative to connection (\ref{C90}) is conserved. It is exactly  the
energy of the massive particle with respect to the reference
frame $\G$.

\section{N\"oether conservation laws}

Let us consider the weak identity (\ref{C40}) when $u$ is a vertical
vector field on the bundle $Y\to X$. It takes the form
\begin{equation}
s^*\bL_{\ol u}L\ap \frac{d}{dx^\la}T^\la, \label{C73}
\end{equation}
$$
(u^i\dr_i +\wh\dr_\la u^i\dr_i^\la)\cL \ap \wh\dr_\la(\pi^\la_i u^i),
$$
where
\begin{equation}
T=u\rfloor\Xi_L=\pi^\la_i u^i\om_\la \label{C74}
\end{equation}
is called the N\"oether current relative to the vertical vector field $u$.

If the Lie derivative of the Lagrangian density $L$ by the vertical
vector field $u$ satisfies the strong condition
$$
\bL_{\ol u}L =0,
$$
we get the weak conservation law (\ref{C42}) of the N\"oether current
(\ref{C74}):
$$
\frac{d}{dx^\la}T^\la\ap 0.
$$

The gauge theory gives the well-known examples of N\"oether conservation laws.

Let $P\to X$
be a principal bundle with a structure
Lie group $G$ which acts freely and transitively on $P$ on the right:
\begin{equation}
r_g : p\mapsto pg, \quad  p\in P,\quad g\in G. \label{1}
\end{equation}

A principal connection $A$ on
the principal bundle $P\to X$ is defined to be a
$G$-equivariant connection on $P$, i.e. $A:P\to J^1P$ with
\[
j^1r_g\circ A= A\circ r_g
\]
for each canonical morphism (\ref{1}).
There is the 1:1 correspondence between the principal connections on a
principal bundle $P\to X$  and the global sections of the quotient
\begin{equation}
C=J^1 P/G\label{68}
\end{equation}
of the jet bundle $J^1 P\to P$ by the first
order jet prolongations of the canonical morphisms (\ref{1}).
We call
$C\to X$ the bundle of principal connections. It is an affine bundle modelled
on the vector bundle
\[
\ol C =T^*X \otimes V^GP
\]
where
\[
 V^GP=VP/G
\]
is the quotient of the vertical tangent bundle $VP$ of $P$
by the canonical action of $G$ on $VP$.
Its standard fiber is the
Lie algebra $\cG_r$ of the right-invariant vector fields on the group
$G$. The group $G$ acts on this standard fiber by the adjoint representation.

Given a bundle atlas $\Psi^P$ of $P$, the  bundle of principal connections $C$
is provided with  the bundle coordinates $(x^\mu,k^m_\mu)$ so that, for every
section $A$ of $C$,
 \[
(k^m_\mu\circ A)(x)=A^m_\mu(x)
\]
are the coefficients of the local connection 1-forms on $X$ corresponding to
the principal connection $A$ with respect to the atlas $\Psi^P$.
The first order jet manifold $J^1C$ of the bundle $C$ is
provided with the adapted coordinates $
(x^\mu, k^m_\mu, k^m_{\mu\la}).$

Let
\[
E=(P\times V)/G
\]
be a vector bundle associated with the principal bundle $P\to X$. Its
sections describe
matter fields.
Every principal connection $A$ on the principal bundle $P$ yields the
associated connection
\begin{equation}
\G_A=dx^\la\otimes [\dr_\la +A^m_\mu (x)I_m{}^i{}_jy^j\dr_i] \label{S4}
\end{equation}
on $E$ where $A^m_\mu (x)$ are the coefficients of the local connection
1-forms and $I_m$ are the generators of the structure group $G$
on the standard fiber $V$ of the bundle $E$.

In case
of unbroken symmetries, the total configuration space of gauge theory
is the product
\begin{equation}
J^1E\op\times_X J^1C. \label{C75}
\end{equation}

In gauge theory, several types of gauge transformations are considered.
To get the N\"oether conservation laws, we restrict our consideration to
vertical isomorphisms of the principal bundle $P$. These are
the $G$-equivariant
isomorphism $\Phi$ of $P$ over $\Id X$, that is,
\begin{equation}
r_g\circ\Phi=\Phi\circ r_g, \qquad g\in G. \label{C93}
\end{equation}
We call them the gauge isomorphisms.
As is well-known, they yield the vertical isomorphisms of the  bundle  of
principal connections  $C$ and the $P$-associated bundle $E$.

Let $u_\cG$ denote a vertical vector field corresponding to a local
1-parameter group of gauge isomorphisms on $P$. There is the 1:1
correspondence between these fields and sections of the bundle $V^GP$. We call
them principal vector field.  The corresponding vector fields on the
$P$-associated vector bundle $E\to X$ which we denote by the same symbol
$u_\cG$ reads \[
u_\cG=\al^m(x)I_m{}^i{}_jy^j\dr_i
\]
where  $\al^m(x)$ are the local components of $u_\cG$ on $P$.
The corresponding vector field on the
bundle of principal connections $C$ takes the form
\begin{equation}
u_\cG=(\dr_\mu\al^m+c^m_{nl}k^l_\mu\al^n)\dr^\mu_m . \label{C85}
\end{equation}
Hence, a principal vector field on the product $C\op\times_XE$ can be
written as
\begin{equation}
u_\cG = (u^{A\la}_m\dr_\la\al^m+ u^A_m\al^m)\dr_A=
 (\dr_\mu\al^m+c^m_{nl}k^l_\mu\al^n)\dr^\mu_m +
\al^m(x)I_m{}^i{}_jy^j\dr_i \label{C76}
\end{equation}
where the collective index $A$ is employed.

A Lagrangian density $L$  on the configuration space  (\ref{C75}) is gauge
invariant iff, for any principal vector field $u_\cG$ (\ref{C76}), we
have the strong equality
$$
{\bf L}_{\ol u_\cG} L=0.
$$

In this case, the first variational formula (\ref{C31}) leads to the strong
equality
$$
(u^A_m\al^m + u^{A\m}_m\dr_\m\al^m)\dl_A\cL +
\wh\dr_\la[(u^A_m\al^m + u^{A\m}_m\dr_\m\al^m)\dr^\la_A\cL]=0
$$
where $\dl_A\cL$ are the variational derivatives of $L$.
Due to arbitrariness of the functions $\al^m(x)$, this equality is equivalent
to the system of the strong equalities
\bea
&& u^A_m\dl_A\cL + \wh\dr_\m(u^A_m\dr^\m_A\cL)=0, \label{D1a}\\
&& u^{A\mu}_m\dl_A\cL
+ \wh\dr_\la(u^{A\mu}_m\dr^\la_A\cL) + u^A_m\dr^\mu_A\cL =0,\label{D1b}\\
&& u^{A\la}_m\dr^\mu_A\cL+ u^{A\mu}_m\dr^\la_A\cL=0. \label{D1c}
\eea
Substituting the equalities  (\ref{D1b}) and (\ref{D1c}) into the equality
(\ref{D1a}), we get the well-known constraint conditions on the variational
derivatives of the gauge invariant Lagrangian density:
$$
u^A_m\dl_A\cL -\wh\dr_\m(u^{A\mu}_m\dl_A\cL)=0.
$$

On solutions of the Euler-Lagrange equations, these equalities come to the
familiar N\"oether identities for a gauge
invariant Lagrangian density $L$:
\bea
&&  \wh\dr_\m(u^A_m\dr^\m_A\cL)\ap 0, \label{D2a}\\
&& \wh\dr_\la(u^{A\mu}_m\dr^\la_A\cL)
+ u^A_m\dr^\mu_A\cL \ap 0,\label{D2b}\\
&& u^{A\la}_m\dr^\mu_A\cL+
u^{A\mu}_m\dr^\la_A\cL=0. \label{D2c}
\eea
A glance at the identity (\ref{D2b}) shows that the current
$$
J^\m_m=u^A_m\dr^\m_A\cL
$$
is reduced to the superpotential. Therefore the identity (\ref{D2a}) is a
consequence of the identities (\ref{D2b}) and (\ref{D2c}).

The weak identities (\ref{D2a}) -- (\ref{D2c}) are the necessary and
sufficient conditions that the weak conservation law
\begin{equation}
\wh\dr_\la[(u^A_m\al^m + u^{A\m}_m\dr_\m\al^m)\dr^\la_A\cL]\ap 0 \label{C300}
\end{equation}
is conserved under the gauge transformations. It means that, if the equality
(\ref{C300}) takes place for a given parameter function $\al(x)$, it remains
true for arbitrary deviations $\al + \dl$ of $\al$. As a consequence,
we observe that,
in virtue of the strong equalities  (\ref{D1b}) and (\ref{D1c}), the
conserved N\"oether current can be brought into form
$$
T^\la=(u^A_m\al^m + u^{A\m}_m\dr_\m\al^m)\dr^\la_A\cL =-\al^m
u^{A\la}_m\dl_A\cL +\wh\dr_\m(\al^m
u^{A\m}_m\dr^\la_A\cL),
$$
and hence it is reduced to the superpotential
$$
T^\la\ap \wh\dr_\m(\al^m u^{A\m}_m\dr^\la_A\cL).
$$
Substituting this superpotential into the conservation law (\ref{C300}), one
comes to the identity (\ref{D2a}) which is the parameter-free form of the
conservation law (\ref{C300}).

Note that another Lepagian equivalent (\ref{C56}) contributes
to the
standard N\"oether currents (\ref{C74})
with the
superpotential term
$$
\wh\dr_\m u^ic_i^{\la\m}\om_\la.
$$

At the same time, one may utilize also the strong identity (\ref{C47}) when
$\e$ is a closed $n$-form on the  bundle of principal connections $C$ which
coresponds to some characteristic class $r$ of the principal bundle $P$ (see
Section 12). Such a form $\e$ is the Lepagian equivalent of the topological
Lagrangian density associated with the corresponding closed characteristic form
$$
r(F)=A^*\e
$$
on $X$ where $F$ is the strenght of the principal connection $A$. Since this
Lagrangian density is gauge invariant, the identity (\ref{C47}) takes the form
$$
0=d_Hh_0(u_\cG\rfloor\e)
$$
and so provides the superpotential term $h_0(u_\cG\rfloor\e)$ in the
N\"oether currents.

\section{General covariance condition}

In this Section, we consider the class of bundles $T\to X$ which admit the
canonical lift of vector fields $\tau$ on $X$. They are called the
bundles of geometric objects. In fact, such canonical lift is the particular
case of the horizontal lift of a field $\tau$ with respect to the suitable
connection on the bundle $T\to X$.

Let $\tau=\tau^\m\dr_\m$ be a vector field on the manifold $X$. There exists
the canonical lift
\begin{equation}
\wt\tau =T\tau= \tau^\m\dr_\m +\dr_\nu\tau^\al\dot x^\nu\frac{\dr}{\dr\dot
x^\al} \label{C91}
 \end{equation}
of $\tau$ onto the tangent bundle $TX$ of $X$.
This lift
consists with the horizontal lift of $\tau$ by means the symmetric connection
$K$ on the tangent bundle which has $\tau$ as the integral section or
as the geodesic field:
$$
\dr_\nu\tau^\al +K^\al{}_{\m\nu}\tau^\m=0.
$$

Generalizing the canonical
lift (\ref{C91}), one can construct the canonical lifts of a vector field
$\tau$ on $X$ onto the following bundles over $X$. For the sake of simplicity,
we denote all these lifts by the same symbol $\wt\tau$.
We have:
\begin{itemize}
\item the canonical lift
$$
\wt\tau = \tau^\m\dr_\m -\dr_\bt\tau^\nu\dot x_\nu\frac{\dr}{\dr\dot x_\bt}
$$
of $\tau$ onto the cotangent bundle $T^*X$;

\item the canonical lift
$$
\wt\tau = \tau^\m\dr_\m + [\dr_\nu\tau^{\al_1}
\dot x^{\nu\al_2\cdots\al_m}_{\bt_1\cdots\bt_k} + \ldots -
\dr_{\bt_1}\tau^\nu \dot x^{\al_1\cdots\al_m}_{\nu\bt_2\cdots\bt_k} -\ldots]
\frac{\dr}{\dr \dot x^{\al_1\cdots\al_m}_{\bt_1\cdots\bt_k}}
$$
of $\tau$ onto the tensor bundle
$$
T^k_mX=(\op\ot^mTX)\ot(\op\ot^kT^*X);
$$

\item the canonical lift
$$
\wt\tau = \tau^\m\dr_\m +[\dr_\nu\tau^\al k^\nu{}_{\bt\m} - \dr_\bt\tau^\nu
k^\al{}_{\nu\m} - \dr_\m\tau^\nu
k^\al{}_{\bt\nu} -\dr_{\bt\m}\tau^\al]\frac{\dr}{\dr k^\al{}_{\bt\m}}
 $$
of $\tau$ onto the bundle $C$ of the linear connections on $TX$.
\end{itemize}

One can think of the vector fields  $\wt\tau$ on a bundle of geometric objects
$T$ as being the vector fields associated with local 1-parameter groups of the
holonomic isomorphisms of $T$ induced by diffeomorphisms of its base $X$. In
particular, if $T=TX$ they are the tangent isomorphisms. We call these
isomorphisms the general covariant transformations.

Let $T$ be the bundle of geometric objects and $L$ a Lagrangian density on
the configuration space $J^1T$. Given a vector field $\tau$ on the base $X$
and its canonical lift $\wt\tau$ onto $T$, one may utilize the first
variational formula (\ref{C45}) in order to get the corresponding SEM
transformation law. The left side of this formula can be simplified if the
Lagrangian density satisfies the general covariance condition.

 Note that, if the Lagrangian density $L$ depends on
background fields, we should consider the corresponding total bundle
(\ref{C41}) and the Lagrangian density $L_{\rm tot}$ on the total
configuration space  $J^1T_{\rm tot}$.
We say that the Lagrangian density $L$
satisfies the general covariance condition if $L_{\rm tot}$ is invariant under
1-parameter groups of general covariant transformations of
$T_{\rm tot}$
induced by diffeomorphisms of the base $X$.
It takes place iff, for any vector
field $\tau$ on $X$, the Lagrangian density $L_{\rm tot}$ obeys the equality
\begin{equation}
\bL_{j^1_0\wt\tau}L_{\rm tot}=0 \label{C94}
\end{equation}
where $\wt\tau$ is the canonical lift of $\tau$ onto $T_{\rm tot}$ and
$j^1_0\wt\tau$ is the jet lift of  $\wt\tau$ onto $J^1T_{\rm tot}$.

If the Lagrangian density $L$ does not depend on background fields, the
equality (\ref{C94}) comes to
\begin{equation}
\bL_{j^1_0\wt\tau}L=0. \label{C95}
\end{equation}
Substituting it in the first variational formula (\ref{C45}), we get the week
conservation law
\begin{equation}
0\ap d_Hh_0(\wt\tau\rfloor\Xi_L). \label{C96}
\end{equation}
One can show that the conserved quantity is reduced to a superpotential term.

Here, we verify this fact in case of a tensor bundle $T\to X$. Let it be
coordinatized by $(x^\la, y^A)$ where the collective index $A$ is employed.
Given a vector field $\tau$ on $X$, its canonical lift $\wt\tau$ on $T$ reads
$$
\wt\tau =\tau^\la\dr_\la + u^A{}_\al^\bt\dr_\bt\tau^\al\dr_A.
$$

Let a  Lagrangian density $L$ on the configuration space $J^1T$ be
invariant under general covarian transformations. Then, it satisfies the
equality (\ref{C95}) which takes the coordinate
form
\begin{equation}
\dr_\al(\tau^\al\cL) + u^A{}_\al^\bt\dr_\bt\tau^\al\dr_A\cL +
\wh\dr_\m(u^A{}_\al^\bt\dr_\bt\tau^\al)\dr_A^\m\cL -
 y^A_\al\dr_\bt\tau^\al\dr_A^\bt\cL =0. \label{C310}
\end{equation}
Due to the arbitrariness of the functions $\tau^\al$, the equality
(\ref{C310}) is equivalent to the system of the equalities
\bea
&& \dr_\la\cL=0, \\
&& \dl^\bt_\al\cL + u^A{}_\al^\bt\dr_A\cL +
\wh\dr_\m(u^A{}_\al^\bt)\dr_A^\m\cL -
 y^A_\al\dr_A^\bt\cL =0,\label{C311a}\\
&& u^A{}_\al^\bt\dr_A^\m\cL +u^A{}_\al^\m\dr_A^\bt\cL =0. \label{C311b}
\eea
It is readily observed that the equality (\ref{C311a}) can be brought into
the form
\begin{equation}
\dl^\bt_\al\cL + u^A{}_\al^\bt\dl_A\cL +
\wh\dr_\m(u^A{}_\al^\bt\dr_A^\m\cL) =
 y^A_\al\dr_A^\bt\cL\label{C312}
\end{equation}
where $\dl_A\cL$ are the variational derivatives of the Lagrangian density
$L$. Substituting the relations (\ref{C312}) and (\ref{C311b})
into the weak identity
$$
0\ap \wh\dr_\la [(u^A{}_\al^\bt\dr_\bt\tau^\al -y^A_\al\tau^\al)\dr^\la_A\cL
+\tau^\la\cL]
$$
(\ref{C96}), we get the conservation law
\begin{equation}
0\ap \wh\dr_\la [-u^A{}_\al^\la\dl_A\cL\tau^\al -
\wh\dr_\m(u^A{}_\al^\la\dr_A^\m\cL\tau^\al)]
\label{C313}
\end{equation}
where the conserved current is reduced to the superpotential term
\begin{equation}
Q_{\wt\tau}{}^\la = -u^A{}_\al^\la\dl_A\cL\tau^\al -
\wh\dr_\m(u^A{}_\al^\la\dr_A^\m\cL\tau^\al).
\label{C314}
\end{equation}

In Part 2 of the work, we shall turn to the equality (\ref{C95}) and the
conservation law (\ref{C96}) in case of the bundle of linear connections.

For general field models, we have the product $T\times Y$ of a
bundle $T\to X$ of geometric objects and some other bundle $Y\to X$. The
lift of a vector field $\tau$ on the base $X$ onto the corresponding
configuration space $$J^1T\op\times_X J^1Y$$ reads
$$
\ol\tau = j^1_0\wt\tau +
\tau^\m\G^i_\m\dr_i +(\dr_\la(\tau^\m\G^i_\m)
+\tau^\m y^j_\la\dr_j\G^i_\m-y^i_\m\dr_\la\tau^\m)\dr^\la_i
$$
where $\G$ is a connection on the bundle $Y\to X$.

In this case, we can not say anything about the general covariance condition
independently on the invariance of a Lagrangian density with respect to the
internal symmetries.

For instance, let $P\to X$ be a principal bundle with the structure Lie group
$G$. Let us consider general gauge isomorphisms $\Phi$ of this principal bundle
over diffeomorphisms of the base $X$. They satisfy the relation (\ref{C93}).
We denote by $u_G$ the projectable vector fields on $P$ corresponding to
local 1-parameter groups of such isomorphisms. There is the 1:1 correspondence
between these vector fields and sections of the bundle
$$
T^GP=TP/G.
$$
We call them the general principal vector fields.
In particular, one can show that, given a vector field $\tau$ on the base $X$,
its horizontal lift onto the principal bundle $P$ by means of a principal
connection on $P$ is a general principal vector field \cite{giach}.

General gauge isomorphisms of the principal bundle $P$, as like as its vertical
isomorphisms, yield the corresponding  isomorphisms of the associated bundles
$E$ and the bundle of principal connections $C$. We denote by the same symbol
$u_G$ the corresponding general principal vector fields on these bundles.

Let
us consider the product
$$
S=C\op\times_X E \op\times_X T
$$
where $T\to X$ is a bundle of geometric objects. Let a Lagrangian density $L$
on the corresponding configuration space $J^1S$ be invariant under the
isomorphisms of the bundle $S$ which are general gauge isomorphisms of
$C\times E$ over diffeomorphisms of the base $X$ and the general
covariant transformations of $T$ induced by these diffeomorphisms
of  $X$.  In particular, vertical isomorphisms of $S$ consist of vertical
isomorphisms of $C\times E$ only. It
should be emphasized that the general gauge isomorphisms of the bundle
$C\times E$ and those of the bundle $T$ taken separately are not the bundle
isomorphisms of the product $S$ because they must cover the
same diffeomorphisms of the base $X$ of
 $Y$. At the same time, one can say that the Lagrangian density $L$
satisfies the general covariance condition in the sense that it is invariant
under general isomorphisms of the bundle $S$ \cite{giach}.

This is phrased in terms of the Lie derivatives as follows. Keeping the
notation of the previous Section, let
$$
u_G = \tau^\la\dr_\la + u^A\dr_A
$$
be a
general principal vector field on the product $C\times E$ which is projected
onto the vector field
$$
\tau=\tau^\la\dr_\la
$$
on the base $X$. The corresponding general principal vector field on the bundle
$Y$ reads
\begin{equation}
\wt u_G= \wt\tau + u^A\dr_A \label{C99}
\end{equation}
where $\wt\tau$ is the canonical lift of $\tau$ onto the bundle of geometric
objects $T$. A Lagrangian density $L$ is invariant under general
isomorphisms of the bundle $S$ iff
\begin{equation}
\bL_{j^1_0\wt u_G}L=0 \label{C98}
\end{equation}
where the jet lift $j^1_0\wt u_G$ of the vector field $\wt u_G$ takes the
coordinate form
$$
j^1_0\wt u_G = j^1_0\wt\tau -y^A_\m\dr_\la u^A\dr_A^\la
+ u^A\dr_A + \wh\dr_\la u^A\dr_A^\la.
$$

There are the topological field theories, besides the gravitation theory,
where we can utilize the condition (\ref{C98}).

\section{SEM tensor of matter fields}

In gauge theory, scalar matter fields possessing internal symmetries are
described by sections of a vector bundle
\[E=(P\times V)/G\]
associated with a principal bundle $P$. This bundle is assumed to be
provided with a $G$-invariant fiber metric $a^E$. Because of the canonical
vertical splitting $VE=E\times E$, the metric $a^E$ is a
fiber metric in the vertical tangent bundle $VE\to X$.
Every principal connection $A$ on the principal bundle $P$ yields the
associated connection (\ref{S4}) on $E$.

On the configuration space $J^1Y$ coordinatized by $(x^\la, y^i, y^i_\la)$, the
Lagrangian density of matter fields in the presence of a background connection
$\G_A$ (\ref{S4}) on $Y$ and a background metric $g$ on $X$ reads
\begin{equation}
L_{(m)}=\frac12a^E_{ij}[g^{\m\n}(y^i_\m-\G^i_\m)
(y^j_\n-\G^j_\n)-m^2y^iy^j]\sqrt{| g|}\om.\label{5.12}
\end{equation}

Let us consider the weak identity (\ref{C52}) when $L$ is the Lagrangian
density (\ref{5.12}) and $\G$ is the connection $\G_A$. It is brought into the
form
\begin{equation}
-t_\bt^\al\sqrt{\mid g\mid}\{^\bt{}_{\m\al}\}+
\G^i_\m\wt\dr_i\cL +(y_\la^j-\G^j_\la)\dr_j\G^i_\m\pi^\la_i+ \pi^\la_i
R^i_{\la\m} + \wh\dr_\la [\pi^\la_i(y_\m^i -\G^i_\m)-\dl^\la_\m\cL]\ap
0  \label{C106}
\end{equation}
where $\wt\dr_i$ denote the partial derivatives with respect to the
coordinates $y^i$ on which the Lagrangian density $L_{(m)}$ depends in the
explicit form, and
$$
R^i_{\la\m}
=F^m_{\la\m}I_m{}^i{}_js^j
$$
is the curvature of the connection $\G_A$.

It is easily to verify that
$$
I_m{}^i{}_jy^j\wt\dr_i\cL=0,
$$
$$
I_m{}^i{}_j(y_\la^j-\G^j_\la)\pi^\la_i=0,
$$
and that, for any matter field $s$,  the SEM tensor relative to the
connection $\G_A$ consists with the metric energy-momentum tensor:
 $$
\pi^\la_i(\dr_\m s^i -\G^i_\m)-\dl^\la_\m\cL=t_\m^\la\sqrt{\mid g\mid}.
$$
Hence, the SEM transformation law (\ref{C106}) for matter fields comes to the
covariant transformation law
\begin{equation}
\sqrt{\mid g\mid }\nabla_\la t^\la_\m \ap -\pi^\la_i
R^i_{\la\m} \label{C111}
\end{equation}
where $\nabla_\al$ denotes the covariant derivative relative to the
Levi-Civita connection $\{^\bt{}_{\m\al}\}$ of the background world metric $g$.

\section{SEM tensors of gauge potentials}

 In contrast with the matter fields, different  gauge potentials require
different SEM tensors.

Let $P\to X$ be a principal bundle with a structure semisimple
Lie group $G$ and $C$ the bundle of principal connections (\ref{68})
coordinatized by $(x^\m, k^m_\m)$.

There exists the canonical splitting
\begin{equation}
J^1C=\ol C_+\op\oplus_C \ol C_-=(J^2P/G)\op\oplus_C
(\op\w^2 T^*X\op\ot_C V^GP), \label{N31}
\end{equation}
\[ k^m_{\m\la}=\frac12(k^m_{\m\la}+k^m_{\la\m}+c^m_{nl}k^n_\la k^l_\m)
+\frac12( k^m_{\m\la}-k^m_{\la\m} -c^m_{nl}k^n_\la k^l_\m),\]
over $C$. There are the corresponding surjections
\be
&&{\cal S}: J^1 C\to \ol C_+, \qquad {\cal S}^m_{\la\m}=
k^m_{\m\la}+k^m_{\la\m} +c^m_{nl}k^n_\la k^l_\m,\\
&& \cF: J^1 C\to \ol C_-,\qquad
\cF^m_{\la\m}= k^m_{\m\la}-k^m_{\la\m} -c^m_{nl}k^n_\la k^l_\m.
\ee
In particular, if $A$ is a section of $C$,  then
$$
F=\cF\circ A
$$
is the strenght of $A$.

On the configuration space (\ref{N31}),
the conventional Yang-Mills Lagrangian density $L_{\rm YM}$ of gauge
potentials in the presence of a background world metric $g$ on the base $X$
is given by the expression
\begin{equation}
L_{\rm YM}=\frac{1}{4\ve^2}a^G_{mn}g^{\la\m}g^{\bt\n}\cF^m_{\la
\beta}\cF^n_{\m\n}\sqrt{| g|}\,\om \label{5.1}
\end{equation}
where  $a^G$ is a nondegenerate $G$-invariant metric
in the Lie algebra of $G$.

Given a symmetric connection $K$ on the tangent bundle $TX$, every
principal connection $B$ on $P$ gives rise to the connection
\begin{equation}
\G^m_{\m\la}= \dr_\m B^m_\la -c^m_{nl}
k^n_\m B^l_\la -K^\bt{}_{\m\la}(B^m_\bt-k^m_\bt) \label{C107}
\end{equation}
on the bundle of principal connections $C$.

Let $\tau$ be a vector field on the base $X$ and
\begin{equation}
\tau_{BK}=\tau^\la\{\dr_\la +[\dr_\m B^m_\la -c^m_{nl}
k^n_\m B^l_\la -K^\bt{}_{\m\la}(B^m_\bt-k^m_\bt)]\dr^\m_m\} \label{C108}
\end{equation}
its
horizontal lift  onto $C$ by means of the connection (\ref{C107}). For every
vector field $\tau$, one can choose the connection $K$ on the tangent bundle
$TX$ which has $\tau$ as the geodesic field. In this case, the horizontal lift
(\ref{C108}) of the vector field $\tau$ comes to its canonical lift
\begin{equation}
\tau_B=\tau^\la\dr_\la +[\tau^\la(\dr_\m B^m_\la -c^m_{nl}
k^n_\m B^l_\la) +\dr_\m \tau^\la(B^m_\la-k^m_\la)]\dr^\m_m \label{C109}
\end{equation}
by means of the principal connection $B$ on the principal bundle $P$
\cite{giach}. The vector field (\ref{C109}) is just the general principal
vector field on $C$ that has been mentioned in the previous Section.
Hence, the Lie
derivative of the Lagrangian density (\ref{5.1}) by the jet lift $\ol \tau_B$
of the field $\tau_B$ comes to
$$
\bL_{\ol \tau_B}L_{\rm YM}=(\dr_\la\tau^\la \cL_{\rm YM}+\tau^\la\dr_\la
\cL_{\rm YM}-
\cF^m_{\m\nu}\dr_\la\tau^\m\pi^{\nu\la}_m)\om.
$$
The corresponding SEM transformation law takes the form
\ben
&&\dr_\la\tau^\la \cL_{\rm YM} -\tau^\m t_\bt^\al\sqrt{\mid
g\mid}\{^\bt{}_{\m\al}\}- \cF^m_{\m\nu}\dr_\la\tau^\m\pi^{\nu\la}_m
\ap \label{C110}\\
&& \qquad \wh\dr_\la[\pi^{\nu\la}_m(\tau^\m(\dr_\nu B^m_\m -c^m_{nl}
k^n_\nu B^l_\m) +\dr_\nu \tau^\m (B^m_\m-k^m_\m)
-\tau^\m k^m_{\nu\m})+\dl^\la_\m\tau^\m\cL_{\rm YM}]\nonumber
\een
where
$$
t_\bt^\al=\frac{1}{\sqrt{\mid
g\mid}}(\pi^{\nu\al}_m\cF^m_{\bt\nu}- \dl^\al_\bt\cL_{\rm YM})
$$
is the metric energy-momentum tensor of gauge potentials.

Let $A$ be a solution of the Yang-Mills equations. Let us consider the lift
(\ref{C109}) of the vector field $\tau$ on $X$ onto $C$ by means of the
principal connection $B=A$.  In this case, the SEM transformation law
(\ref{C110}) on the critical section $A$ takes the form
$$
\tau^\m t^\al_\bt\sqrt{\mid g\mid }\{^\bt{}_{\m\al}\}\ap \tau^\m
\frac{d}{dx^\la} (\pi^{\nu\la}_m F^m_{\m\nu}- \dl^\la_\m\cL_{\rm YM}),
$$
 and thus it comes
to the covariant
conservation law
\begin{equation}
\sqrt{\mid g\mid }\nabla_\la t^\la_\m\ap 0.\label{C112}
\end{equation}

Note that, in general case of the principal connection $B$, the
corresponding SEM transformation law (\ref{C110}) differs from the covariant
conservation law (\ref{C112}) in the N\"oether conservation law
$$
\wh\dr_\la(\pi^{\nu\la}_m u_{\cG}{}^m_\nu)\ap 0
$$
where
$$
u_\cG=(\dr_\nu\al^m+c^m_{nl}k^l_\nu\al^n)\dr^\nu_m,
$$
$$
\al^m=\tau^\m(B^m_\mu -A^m_\mu)
$$
is the principal vector field (\ref{C85}) on $C$.

It should be emphasized that, in order to get the energy-momentum
transformation laws (\ref{C111}) and (\ref{C112}), the gauge symmetries of
the Lagrangian densities (\ref{5.12}) and (\ref{5.1}) respectively have been
used.

\section{SEM tensors of Proca fields}

Proca fields which are described by sections of the
cotangent bundle $T=T^*X$ exemplify a field model on bundles of geometric
objects, without internal symmetries. In Part 2 of the work, we shall utilize
Proca fields as the matter source of a gravitational field. In this Section,
the SEM transformation law of Proca fields in the presence of a background
world metric is examined.

The
configuration space
$J^1T$
of Proca fields is coordinatized by
$(x^\la,k_\mu,k_{\mu\la})$
where $k_\mu=\dot x_\mu$ are the familiar induced coordinates of $T^*X$.
On this configuration space,
the Lagrangian density of Proca fields is written as
\begin{equation}
L_{\rm P}=[-\frac{1}{4\g}g^{\mu\al}g^{\nu\beta}\cF_{\al
\beta}\cF_{\mu\nu}
 -\frac12 m^2g^{\mu\la}k_\mu k_\la]\sqrt{\mid g\mid}\omega \label{214}
\end{equation}
where
$$
\cF_{\mu\nu}=k_{\nu\m} -k_{\mu\nu}.
$$
The associated Poincar\'e-Cartan form on $J^1T$ reads
\begin{equation}
\Xi_{\rm P}= (\cL_{\rm P}
 -\pi^{\m\la}k_{\mu\la})\om + \pi^{\m\la}dk_\mu\w\om_\la,
\label{C301}
\end{equation}
$$
\pi^{\m\la}= -\frac1{\g}g^{\mu\al}g^{\la\beta}\cF_{\beta\al}\sqrt{\mid g\mid}.
$$
The Euler-Lagrange operator is
\begin{equation}
\cE_{\rm P}= \sqrt{\mid g\mid}[- m^2g^{\al\bt}k_\al -\wh\dr_\m\pi^{\bt\m}
]dk_\bt\w\om=\dl^\bt\cL dk_\bt\w\om
\label{C303}
\end{equation}
where $\dl^\bt\cL $ are variational derivatives of the Lagrangian density
$L_{\rm P}$.

Let $\tau$ be a vector field on the base $X$ and
$$
\wt\tau = \tau^\m\dr_\m -\dr_\al\tau^\nu k_\nu\frac{\dr}{\dr k_\al}
$$
its canonical lift onto $T^*X$.

The Lie
derivative of the Lagrangian density (\ref{214}) by the jet lift
$j^1_0\wt\tau$
of the field $\wt\tau$ is
$$
\bL_{j^1_0\wt\tau}L_{\rm P}=(\dr_\la\tau^\la \cL_{\rm P}+\tau^\la\dr_\la
\cL_{\rm P}-
\cF_{\m\nu}\dr_\la\tau^\m\pi^{\nu\la} + m^2g^{\nu\la}\dr_\la\tau^\m
k_\nu k_\m\sqrt{\mid g\mid})\om.
 $$
Then, the corresponding SEM transformation law
$$
\bL_{j^1_0\wt\tau}L_{\rm P}\ap \wh\dr_\la[\pi^{\nu\la}(-\dr_\nu\tau^\m k_\m
-\tau^\m k_{\nu\m}) +\tau^\la\cL]
 $$
(\ref{J4}) takes the form
\begin{equation}
-\dr_\la \tau^\m t_\m^\la\sqrt{\mid g\mid}-\tau^\m t_\bt^\al\sqrt{\mid
g\mid}\{^\bt{}_{\m\al}\} \ap
 \wh\dr_\la[-\tau^\m t_\m^\la\sqrt{\mid g\mid}+ \tau^\nu k_\nu\dl^\la\cL
-\wh\dr_\m(\pi^{\mu\la}\tau^\nu k_\nu)], \label{C305}
\end{equation}
\begin{equation}
-t_\mu^\la\sqrt{\mid g\mid}=
\pi^{\nu\la}\cF_{\nu\m} + m^2g^{\nu\la}
k_\mu k_\nu\sqrt{\mid g\mid}+\dl^\la_\m \cL_{\rm P}. \label{C304}
\end{equation}

A glance at the expression (\ref{C305}) shows that the SEM tensor of the
Proca field
\begin{equation}
\cT_{\ol\tau}{}^\la =\tau^\m t_\m^\la\sqrt{\mid g\mid}- \tau^\nu
k_\nu\dl^\la\cL +\wh\dr_\m(\pi^{\mu\la}\tau^\nu k_\nu) \label{307}
\end{equation}
is the sum of the familiar metric energy-momentum tensor
 and the superpotential term
\begin{equation}
Q_{\ol\tau}{}^\la = - \tau^\nu k_\nu\dl^\la\cL
+\wh\dr_\m(\pi^{\mu\la}\tau^\nu k_\nu) \label{C306}
\end{equation}
which is the particular case of the superpotential term (\ref{C314}).
This term however does not make any contribution into the differential
conservation law (\ref{C305}) which thus takes the standard form
$$
\nabla_\la t^\la_\m\sqrt{\mid g\mid }\ap 0.
$$

At the same time, the superpotential term reflects the partial invariance of
the Lagrangian density (\ref{214}) under general covariant transformations
 broken by the background metric field. In gravitation theory, when
the general covariant transformations are exact,  the total superpotential
term contains
the whole SEM tensor
(\ref{307}) of Proca fields (see Section 14). Thus,
the Proca field model examplifies the phenomenon of "hidden energy". Only the
superpotential part of energy-momentum is displayed if the general covariant
transformations are exact.

It is readily observed that the superpotential term arise due to the fact
that the canonical lift $\wt\tau$ of a vector field $\tau$ on $X$ depends
on the derivatives of the components of the field $\tau$. Therefore,
superpotential terms are the standard attributes of SEM tensors in field
models on bundles of geometric objects.

\section{Topological gauge theories}

The field  models that we have investigated in the fomer Sections
 show that when a background world metric is present, the
SEM transformation law comes to the covariant conservation law of
the metric energy-momentum tensor. Topological gauge theories exemplify the
field models in the absence of a world metric.

Let us consider the Chern-Simons gauge model on a 3-dimensional base manifold
$X^3$ \cite{bir,kha,wit}.

Let $P\to X^3$ be a principal bundle with a structure semisimple Lie group
$G$ and $C$ the corresponding bundle of principal connections (\ref{68})
which is coordinatized by $(x^\la, k^m_\la)$. Keeping the notations of the
previous Section, the Chern-Simons Lagrangian density is given by the
coordinate expression
\begin{equation}
L_{\rm CS} =\frac{1}{2k}a^G_{mn}\ve^{\al\la\m} k^m_\al (\cF^n_{\la\m}
+\frac13c^n_{pq}k^p_\la k^q_\m)d^3x \label{C113}
\end{equation}
where $\ve^{\al\la\m}$ is the skew-symmetric Levi-Civita tensor.

It is readily observed that the Lagrangian density (\ref{C113}) is not gauge
invariant and globally defined. At the same time, it provides the globally
defined Euler-Lagrange operator
$$
\cE_{L_{\rm CS}} =\frac1ka^G_{mn}\ve^{\al\la\m}\cF^n_{\la\m} dk^m_\al\om.
$$
Thus, the gauge transformations in the Chern-Simons model appear to be the
generalized invariant transformations which keep invariant the
Euler-Lagrange equations, but not the Lagrangian density. Solutions of these
equations are the curvature-free principal connections $A$ on the principal
bundle $P\to X^3$.

Though the Chern-Simons Lagrangian density is not invariant under gauge
transformations, we still have the N\"oether-type conservation law in which the
total conserved current is the standard  N\"oether current (\ref{C74}) plus the
additional term as follows.

Let $u_\cG$ be the principal vector field (\ref{C85}) on the bundle of
principal connections $C$. We compute
$$
\bL_{\ol u_\cG}L_{\rm
CS}=\frac1ka^G_{mn}\ve^{\al\la\m}\dr_\al(\al^m\dr_\la A^n_\m)d^3x.
$$
Hence, the N\"oether transformation law (\ref{C73}) comes to the conservation
law
\begin{equation}
\frac{d}{dx^\la} T_{\rm CS}{}^\la =
\frac{d}{dx^\la} (T^\la +\frac1ka^G_{mn}\ve^{\al\la\m}\al^m\dr_\al A^n_\m)
\ap 0 \label{C114}
\end{equation}
where
$$
T^\la = \pi^{\m\la}_nu_\cG{}^n_\m=\frac1ka^G_{mn}\ve^{\al\la\m}A^m_\al
u_\cG{}^n_\m $$
is the standard N\"oether current. After simplification, the
conservation law (\ref{C114}) takes the form
$$
\frac{d}{dx^\la} (\frac1ka^G_{mn}\ve^{\al\la\m}\al^mF^n_{\al\m}) \ap 0.
$$
In the Chern-Simons model, the total conserved current $T_{\rm CS}$ is equal
to zero. At the same time, if we add the Chern-Simons Lagrangian density to the
Yang-Mills one, $T_{\rm CS}$ plays the role of the massive term and makes the
contribution into the standard N\"oether current of the Yang-Mills gauge
theory.

Turn now to the SEM transformation law in the Chern-Simons model.

Let $\tau$ be a vector field on the base $X$ and $\tau_B$ its lift
(\ref{C109}) onto the bundle $C$ by means of a section $B$ of $C$. Remind that
the vector fields $\tau_B$ are the general principal vector fields
associated with local 1-parameter groups of general gauge isomorphisms of
$C$. We compute
$$
\bL_{\ol \tau_B}L_{\rm
CS}=\frac1ka^G_{mn}\ve^{\al\la\m}\dr_\al(\tau^\nu B^m_\nu\dr_\la A^n_\m)d^3x.
$$
The corresponding SEM transformation law takes the form
\begin{equation}
\frac{d}{dx^\la} \cT_{\rm CS}{}^\la =
\frac{d}{dx^\la} (\cT^\la -\frac{1}{2k}a^G_{mn}\ve^{\al\la\m}\tau^\nu
B^m_\nu\dr_\al A^n_\m) \ap 0 \label{C115}
\end{equation}
where
$$
\cT^\la = \pi^{\m\la}_n[\tau^\nu\dr_\nu A^m_\m-\tau^\nu(\dr_\m B^n_\nu
-c^n_{pq} A^p_\mu B^q_\nu) -\dr_\mu \tau^\nu (B^n_\nu-A^n_\nu)
]-\dl^\la_\nu\tau^\nu\cL_{\rm CS}
$$
is the standard SEM tensor relative to the lift $\tau_B$ of the vector field
$\tau$.

Let $A$ be a critical section. We consider the lift (\ref{C109}) of the
vector field $\tau$ on $X$ onto $C$ by means of the principal connection
$B=A$, just as we have done in the previous Section. Then, the SEM conservation
law (\ref{C115}) comes to the conservation law
\begin{equation}
\frac{d}{dx^\la}[\frac1ka^G_{mn}\tau^\nu\ve^{\al\la\m}A^m_\al\cF^n_{\nu\m}
-\tau^\la \cL_{\rm CS}]=
\frac{d}{dx^\la}(-\frac{1}{6k}\tau^\la\ve^{\al\nu\m}c_{npq}A^n_\al A^p_\nu
A^q_\m) \ap 0. \label{C116}
\end{equation}

Note that, since the gauge symmetry of the Chern-Simons Lagrangian density is
broken, the SEM conservation law (\ref{C116}) fails to be invariant under
gauge transformations.

Let us consider Lagrangian densities of topological gauge models which are
invariant under the general gauge isomorphisms of the bundle $C$. Though they
imply the zero Euler-Lagrange operators, the corresponding strong identities
may be utilized as the superpotential terms when such a topological Lagrangian
density is added to the Yang-Mills one.

Let $P\to X$ be a principal bundle with the structure Lie group $G$. Let us
consider the bundle $J^1P\to C$. This also is a G principal bundle. Due to
the canonical vertical splitting
$$
VP=P\times \cG_l
$$
where $\cG_l$ is the left Lie algebra of the group $G$,
the complementary morphism (\ref{24})
of $J^1P$ defines the canonical $\cG$-valued 1-form $\th$ on $J^1P$. This
form is the connection form of the canonical principal
connection on the principal bundle $J^1P\to C$ \cite{gar}. Moreover, if
$$
\G_A: P\to J^1P
$$
is a principal connection on $P$ and $A$ the corresponding connection form, we
have
 $$
 \G_A^*\th =A.
$$
If $\Om$ and $R_A$ are the curvature 2-forms of the connections $\th$ and $A$
respectively, then
$$
\G_A^*\Om =R_A.
$$

Local connection 1-forms on $C$ associated with the canonical connection
$\th$ are given by the coordinate expressions
$$
k^m_\m dx^\m\otimes I_m.
$$
The corresponding curvature 2-form on $C$ reads
$$
\Om_C= (dk^m_\m\w dx^\m -\frac12c^m_{nl}k^n_\m k^l_\nu dx^\nu\w dx^\m)
\otimes I_m.
$$

Let $I(\cG)$ be the algebra of real $G$-invariant polinomial on the Lie
algebra $\cG$ of the group $G$. Then, there is the well-known Weyl homomorphism
of  $I(\cG)$ into the De Rham cohomology algebra $H^*(C,\R)$. In virtue of
this isomorphism, every $k$-linear element $r\in I(\cG)$ is represented by
the cohomology class of the closed characteristic  $2k$-form $r(\Om_C)$ on
$C$. If $A$ is a section of $C$, we have
$$
A^*r(\Om_C)= r(F)
$$
where $F$ is the strenght of $A$ and $r(F)$ is the corresponding characteristic
form on $X$.

 Let $\dim X$ be
even and a characteristic $n$-form $r(\Om_C)$ on $C$ exist. This is a
Lepagian form which defines a gauge invariant Lagrangian density
$$
L_r =h_0(r(\Om_C))
$$
on the jet manifold $J^1C$. The Euler-Lagrange operator associated with $L_r$
is equal to zero. Then, for any projectable vector field $u$ on $C$, we have
the strong relation (\ref{C47}):
$$
\bL_{\ol u}h_0(r(\Om_C)) = h_0(d\ol u\rfloor r(\Om_C)).
$$
If $u$ is a general principal vector field on $C$, this relation takes the
form
$$
0 = d_H(\ol u\rfloor r(\Om_C)).
$$

For instance, let  $\dim X=4$ and the group $G$ be semisimple. Then,  the
 characteristic Chern-Pontriagin 4-form
$$
r(\Om_C)= a^G_{mn}\Om^n_C\w \Om^m_C.
$$
It is the Lepagian equivalent of the Chern-Pontriagin Lagrangian density
$$
L=\frac1ka^G_{mn}\ve^{\al\bt\m\nu}\cF^n_{\al\bt}\cF^m_{\m\nu}d^4x
$$
of the topological Yang-Mills theory.

\end{document}